\documentclass[12pt]{article}
\usepackage{epsfig,amssymb,amsmath,psfrag}
\usepackage{color,colordvi}
\usepackage{ulem}

\catcode`\@=11
\def \Re {\mathop{\rm Re}\nolimits}

\def \e  {\mathop{\rm e}\nolimits}
\newcommand\lr[1]{{\left({#1}\right)}}

\newcommand \ket [1] {|{#1}\rangle}
\newcommand \bra [1] {\langle {#1}|}
\newcommand\re[1]{(\ref{#1})}

\newcommand{\cN}{{\cal N}}

\newcommand{\pa}{\partial}

\renewcommand{\a}{\alpha}
\renewcommand{\b}{\beta}

\def \la {\langle}
\def \ra {\rangle}

\def \eps {\epsilon}

\newcommand{\p}[1]{(\ref{#1})}

\newcommand \vev [1] {\langle{#1}\rangle}
\newcommand{\ft}[2]{{\textstyle\frac{#1}{#2}}}

\def\numberbysection{\@addtoreset{equation}{section}
                     \def\theequation{\thesection.\arabic{equation}}}

\numberbysection

\definecolor{orange}{rgb}{0.5,0.8,0.3}

\begin{document}

\begin{titlepage}

\thispagestyle{empty}

\null\vskip-43pt \hfill
\begin{minipage}[t]{45mm}
CERN-PH-TH/2013-211\\
IPhT--T13--210\\
LAPTH-047/13
\end{minipage}

\vspace*{10mm}

\centerline{\large \bf From correlation functions to event shapes}
\vspace*{9mm}

\centerline{\sc A.V.~Belitsky$^{a,b}$,  S.~Hohenegger$^c$, G.P.~Korchemsky$^b$, E.~Sokatchev$^{c,d,e}$, A.~Zhiboedov$^f$}

\vspace{5mm}

\centerline{\it $^a$Department of Physics, Arizona State University}
\centerline{\it Tempe, AZ 85287-1504, USA}

\vspace{3mm}

\centerline{\it $^b$Institut de Physique Th\'eorique\footnote{Unit\'e de Recherche Associ\'ee au CNRS URA 2306}, CEA Saclay}
\centerline{\it 91191 Gif-sur-Yvette Cedex, France}

\vspace{3mm}
\centerline{\it $^c$Physics Department, Theory Unit, CERN}
\centerline{\it CH -1211, Geneva 23, Switzerland}

\vspace{3mm}
\centerline{\it $^d$Institut Universitaire de France}
\centerline{\it  103, bd Saint-Michel
F-75005 Paris, France }

\vspace{3mm}
\centerline{\it $^e$LAPTH\,\footnote[2]{UMR 5108 du CNRS, associ\'ee \`a l'Universit\'e de Savoie},   Universit\'{e} de Savoie, CNRS}
\centerline{\it  B.P. 110,  F-74941 Annecy-le-Vieux, France}

\vspace{3mm}
\centerline{\it $^f$Department of Physics, Princeton University}
\centerline{\it Princeton, NJ 08544, USA}

\vspace{1cm}

\centerline{\bf Abstract}

\vspace{2mm}

We present a new approach to computing event shape distributions or, more precisely, charge flow correlations in a generic conformal field theory (CFT). These   infrared finite observables are 
familiar from collider physics studies and describe the angular distribution of global charges in outgoing radiation created from the vacuum by some source. The charge flow correlations can be
expressed in terms of Wightman correlation functions in a certain limit. We explain how to compute these quantities starting from their Euclidean analogues by means of a non-trivial analytic
continuation which, in the framework of CFT, can be performed elegantly in Mellin space. The relation between the charge flow correlations and Euclidean correlation functions can be 
reformulated directly in configuration space, bypassing the Mellin representation, as a certain Lorentzian double discontinuity of the correlation function integrated along the cuts. We illustrate 
the general formalism in ${\cal N} = 4$ SYM, making use of the well-known results on the four-point correlation function of half-BPS scalar operators.  We compute the double scalar flow 
correlation in ${\cal N} = 4$ SYM, at weak and strong coupling and show that it agrees with known results obtained by different techniques. One of the remarkable features of the $\cN=4$ theory is 
that the scalar and energy flow correlations are proportional to each other.  Imposing natural physical conditions on the energy flow correlations (finiteness, positivity and regularity), we formulate 
additional  constraints on the four-point correlation functions in ${\cal N} = 4$ SYM that should be valid at any coupling and away from the planar limit. 

\end{titlepage}

\setcounter{footnote} 0

\newpage

\pagestyle{plain}
\setcounter{page} 1

\tableofcontents

\newpage

\section{Introduction}

In this paper we study a class of observables in conformal field theories (CFT) that is {familiar from collider physics} and is known as {\it event shapes} or,
more specifically, {{charge flow correlations}}. The physical picture behind these observables is very general and simple: we excite the vacuum with a probe and study
the produced state with calorimeters {that measure the flow of conserved charges, be it energy or any global symmetry charge, }in a given direction at spatial infinity.

These observables were first introduced in the analysis of $e^+ e^-$ annihilation into hadrons in the context of QCD (for a review see, e.g., 
\cite{Kunszt:1989km,Biebel:2001dm}). In this process, the
electron and positron annihilate to produce a virtual photon $\gamma^*(q)$ with large invariant mass $q^2$. It excites the QCD vacuum and produces quarks
and gluons, which then propagate into the final state and undergo a transition into hadrons. Investigating the distribution of the outgoing particles and their quantum numbers 
(charges, energy, etc.) in the final state, we can obtain detailed knowledge about the underlying QCD dynamics \cite{Sterman:1977wj,Basham:1978zq}. The most 
prominent and best understood observables in this
context are the so-called event shapes or weighted cross sections \cite{Biebel:2001dm}. They are given by (an infinite) sum over the final hadronic states,
\begin{align}\label{event-shapes}
\sigma_w(q) = \sum_X (2\pi)^4 \delta^{(4)}(q-k_X)w(X)|\vev{X|O(0)|0}|^2 \,,
\end{align}
where $\vev{X|O(0)|0}$ describes the creation out of the vacuum of a state $\ket{X}$ with total momentum $k_X$ by a local operator $O(0)$ (here, the
electromagnetic QCD current). The weight factor $w(X)$ depends on the quantum numbers of the final states which one selects in the detector apparatus. Various
event shapes (e.g., thrust, heavy mass, energy-energy correlations) correspond to different choices of  $w(X)$. In the simplest case   $w(X)=1$ we obtain the
total cross section $ \sigma_{\rm tot}(q)$.

Making use of the completeness condition for the hadronic states, $\sum_X \ket{X}\bra{X}=1$, we can re-express the total cross section as a Fourier transform of the 
non-time ordered (Wightman) two-point correlation function\footnote{Using the optical theorem,  the total cross section can also be rewritten as the imaginary part of the 
Fourier transformed time-ordered two-point correlation function of the $O$'s.} of  the $O$'s. An analogous representation in terms of correlation functions 
\cite{Sveshnikov:1995vi,Korchemsky:1997sy,Korchemsky:1999kt,BelKorSte01, Hofman:2008ar} also exists for the charge flow correlations \cite{Basham:1978zq}. 
The very fact that such a reformulation exists follows from  basic properties of quantum field theory, independently of
any dynamical details. Indeed, in any quantum field theory the energy or charge of a state can be measured by  integrating the corresponding conserved currents 
over the spatial volume. Analogously, to measure the flow of a charge at infinity, we insert the conserved current at spatial infinity and integrate it
over the time interval during which the measurement is performed \cite{Korchemsky:1997sy}. In particular, this definition is applicable to CFTs \cite{Hofman:2008ar} 
where the notion of asymptotic states
is ill-defined and formulas like Eq.\ \re{event-shapes} should be interpreted with great care.\footnote{Notice that in perturbation theory one can  define scattering amplitudes within
a given infrared regularization scheme.}

The main subject of this paper is a particular class of event shape distributions, the so-called charge  flow correlations. They can be reformulated in terms of correlation functions
\cite{Sveshnikov:1995vi,Korchemsky:1997sy,Korchemsky:1999kt,BelKorSte01, Hofman:2008ar},
\begin{align}\label{Wightman}
\sigma_w(q) =  \int d^4 x\, \e^{iqx}   \vev{0|  O^\dagger (x) {\mathcal D}[w] O(0)|0} \, ,
\end{align}
where $O^{\dagger}(x)$ stands for the Hermitian conjugate of $O(x)$.
A few comments concerning this expression are in order. Here the operator $O(x)$ (called the `source') creates a state that we are probing. The flow operator $\mathcal{D}[w]$
is determined by a set of local operators, e.g., the stress-energy tensor $T_{\mu\nu}$ or conserved current $J_\mu$,  depending on the choice
of the detectors, collectively called `charges'. In what follows we cumulatively denote these as $\widetilde{O}(x)$ to distinguish them from the source
$O(x)$. The operators $\widetilde{O}(x)$ are inserted at spatial infinity, each pointing in a different direction on the infinite sphere,  and are interpreted as detectors (`calorimeters')
measuring the flow of the corresponding `charge' in a given direction. Finally, the detectors are integrated over the measurement time interval.  The precise definition of
$\mathcal{D}[w]$ as well as its properties will be explained in great detail in the main body of the paper. Another important property of \re{Wightman} is that it is defined by an
intrinsically Lorentzian, non-time ordered (Wightman) correlation function of local operators, e.g., $\la 0| O^\dagger(x) T_{\mu \nu}(x_1) J_{\rho}(x_2) \dots O(0)|0 \ra$. 
Re-expressed in terms of correlation functions of local operators, the event shapes \re{Wightman} are thus defined non-perturbatively at arbitrary coupling in any quantum field 
theory (including CFTs \cite{Hofman:2008ar}).

The relation between the two seemingly different representations \re{event-shapes} and \re{Wightman} can be easily understood in theories that admit an S-matrix. In 
this case we can insert a
complete set of asymptotic states, $\sum_X \ket{X}\bra{X}=1$, labeled by the momenta and charges of the particles, in Eq.\ \re{Wightman}. This basis can 
be chosen to diagonalize the flow
operator, ${\mathcal D}[w] \ket{X}= w(X) \ket{X}$, which brings us back to Eq.\ \re{event-shapes}. In CFTs this operation is, in
general, not well defined due to the absence of asymptotic on-shell states. However, perturbatively one can easily define an S-matrix and apply the same argument 
in a CFT by slightly deforming it
away from the critical point.

The goal of this paper is to further develop the approach to event shapes based on Wightman correlation functions in CFTs. We assume that the
Euclidean versions of the correlators are known and present a step-by-step procedure for computing the charge flow correlations \re{Wightman}. In doing so, we have to analytically
continue the Euclidean correlation functions to their Wightman counterparts \cite{LusMac75} in Minkowski space-time, and then insert them into the definition \re{Wightman}. For 
our purposes, a very efficient way to carry out the required analytic continuation is through the Mellin space representation of the correlation functions \cite{Mack:2009mi}.

The outcome of our analysis is that the charge flow correlations $\sigma_w(q)$ are given by a convolution of the Mellin amplitude, defining the Euclidean correlation 
function, with the so-called detector kernel which is uniquely
fixed by the form of the flow operator $\mathcal D[w]$. This result is completely general since it only relies
on the conformal symmetry of the theory and thus holds for any CFT.\footnote{The same method should be applicable to hypothetical scale-invariant but non-conformal
theories or any other theory where radiation (energy flux) is carried away solely to the future null infinity. For theories with massive particles the detector limit used in 
this paper should be reconsidered.}  We also demonstrate that
the charge flow correlations admit another equivalent representation in terms of a certain Lorentzian double discontinuity of the four-point correlation function convoluted 
with the detector kernel in the coordinate space. Applied to the perturbative expansion of the correlation functions available in the literature, the formalism developed here 
allows us to easily reproduce the known results on the event shapes observables, obtained from \re{event-shapes} using standard techniques.

To appreciate the power of the correlation function approach, it is instructive to compare it with the conventional one based on amplitudes.  Namely,
applying the relation \re{event-shapes}, we can compute the weighted cross section $\sigma_w(q)$ from the transition amplitudes $\vev{X|O(0)|0}$
evaluated  in weakly coupled gauge theories. The close examination of \re{event-shapes} reveals, however, that such an approach has the following disadvantages:
\begin{itemize}
\item presence of intrinsic infrared (IR) divergences;
\item integration over the phase space of the final state and subsequent intricate IR cancellations;
\item necessity for summation over all final states.
\end{itemize}

Let us comment on each of these points. They
are very well understood in the context of perturbation theory\footnote{Here we have in mind some perturbative conformal supersymmetric gauge theory like ${\cal N}=4$ SYM.}
and are treated in textbooks (see, e.g., \cite{StermanBook}). The scattering amplitudes used to compute the
event shapes involve massless particles and   contain IR divergences that require a regularization procedure. The IR singularities are
known to cancel according to the famous Kinoshita-Lee-Nauenberg theorem \cite{Kinoshita:1962ur, Lee:1964is}. However, the cancelation is quite intricate in
nature since it relies on the compensation between the singularities in the virtual corrections and those in the emission of real quanta. Finally, to
compute the weighted cross sections \re{event-shapes}, we have to sum over an infinite set of final states, in which the total number of produced
particles grows linearly with the number of loops. The resulting phase space integrals are known to be extremely hard to handle and so far little progress
has been made in understanding their general  analytic structure.

It is now straightforward to see the advantages of  computing charge flow correlations using the correlation functions \re{Wightman}.
They are summarized in the following three `no' statements:
\begin{itemize}
\item no IR divergences are present in the correlation functions;
\item no summation over all final states is needed;
\item no integration over the phase space is required.
\end{itemize}
There is however a price to pay. In order to apply \re{Wightman}, we have to find an efficient way of computing the Euclidean correlation functions involving the 
product of source operators, $O(0)$
and $O^\dagger (x)$, and several charge operators $\widetilde O(x_i)$, then analytically continue them to Minkowski signature to obtain the corresponding 
Wightman counterparts and, finally, integrate
over the detector (or calorimeter) working time to get $\vev{0|  O^\dagger(x) {\mathcal D}[w] O(0)|0}$. Conformal symmetry simplifies the form of the correlation functions enough to
make this program of computing weighted cross sections from correlation functions an attractive enterprise.

What is the  utility of exploring charge flow correlations in CFTs? There are two immediate applications. The first one comes from the similarity of charge flow 
correlations in certain four-dimensional superconformal theories (e.g., ${\cal N}=4$ SYM) at weak coupling with those in perturbative QCD. The hope is that by 
studying the former one can get some  insight into the latter.  Secondly, in the special case of energy correlations, $\sigma_w(q)$ is expected to be a positive 
definite\footnote{Since the energy-flow observables are intrinsically Lorentzian in nature and probe physics on the light cone, it is not known how to reproduce 
them in Euclidean space. Recent examples where Lorentzian positivity was exploited include Refs.~\cite{Komargodski:2011vj, Luty:2012ww, Komargodski:2012ek}.} 
and  finite function in unitary CFTs \cite{Hofman:2008ar, Hofman:2009ug}. In the case of interacting CFTs we also expect it to be a regular function of the angular variables. 
These conditions impose nontrivial constraints on the correlation functions  of the stress-energy tensor in a generic CFT. Finally, recall that CFTs are ubiquitous in the 
body of modern theoretical physics,  some notable examples being physics of critical phenomena, end-points of renormalization
group flow, quantum gravity in AdS. Therefore, understanding their general properties is of great interest.

In this work we consider flow operators and sources that are built from scalar primary operators. Such operators are not related to conserved Noether 
currents and, as a consequence, their integrals
over the spatial volume do not generate conserved charges.
So, strictly speaking, the scalar flow operators would not produce observables that have a well-defined interpretation as
charge flow correlations. However, this simple example allows us to reveal all of the salient features of our approach. We may say that  the scalar flow operators are 
simplified prototypes
of the more realistic charge flow operators without the complication of dealing with Lorentz indices.   Moreover, in theories with supersymmetry certain scalar primary operators 
are members of 
supermultiplets that contain, e.g.,  the
R-symmetry currents and the stress-energy tensor, which do generate well-defined physical observables. The relation between scalar flow operators and charge flow operators 
built from conserved
currents, in the context of $\cN=4$ SYM,  is explored in the companion paper \cite{paper2}.

Our subsequent presentation is organized as follows. In Section 2, we describe  the relation between charge flow correlations and Euclidean correlation functions. 
In Section 3, we use the
simple example of a free scalar field to show the equivalence of the two representations \re{event-shapes} and \re{Wightman}. In Section \ref{illustration}, we illustrate the
correlation function approach to \re{Wightman} by computing the charge flow correlations in ${\cal N}=4$ SYM with the initial state created by the simplest half-BPS scalar 
operator belonging to the
${\cal N}=4$ stress-tensor multiplet. The same half-BPS operators are used to generate the scalar flows. We compute the scalar flow correlations at weak coupling using the two representations
\re{event-shapes} and \re{Wightman}, and at strong coupling using the available
results on the correlation functions through AdS/CFT \cite{Maldacena:1997re,Gubser:1998bc,Witten:1998qj}. We also present constraints on the correlation function that follow 
from the IR finiteness, regularity and positivity of the energy correlations. Section 5 contains concluding remarks. Some technical details of the calculations are summarized in 
several appendices.

\section{From Euclid to Wightman: general technology}
\label{EuclToWight}

In this section we initiate a thorough exposition of the formalism by outlining all steps involved in the computation of the charge flow observables \re{Wightman} via Wightman
correlation functions by means of analytic continuation of their Euclidean counterparts.
Our discussion is valid for generic CFTs since it relies solely on the conformal symmetry and general properties of  correlation functions.

For the sake of simplicity and transparency we use scalar primary operators to define both the source $O(x)$ and the flow operators $\mathcal D[w]$ in \re{Wightman}, though they will
differ in the choice of quantum numbers, as we explain below.  We opt to work in a
four-dimensional CFT and choose the scaling dimension of the scalar operators in question to be $\Delta = 2$. The reason is that the same value of $\Delta$ corresponds to the
twist (dimension minus spin) of conserved currents in $D = 4$. In principle, we could have chosen both the space-time dimension $D$ and the scaling dimension of the 
scalar operator $\Delta$ to be arbitrary. In Appendix \ref{DdimCFT} we show that the resulting expression for the weighted cross section \re{Wightman} is more involved 
in this general case, but it simplifies significantly for the aforementioned values $\Delta=2$ and $D=4$.

In supersymmetric theories composite operators are organized into supermultiplets. If the latter contain at least one conserved current, 
all  members of the multiplet have  zero anomalous
dimensions. So, we can build a detector from any component of the multiplet including those
which do not generate conserved charges. Indeed, this is what
happens, for example, in the ${\cal N}=4$ Yang-Mills superconformal field theory (SCFT). There, the scalar  half-BPS operators $O(x)$ with the 
scaling dimension $\Delta=2$ are the lowest components
of an $\mathcal{N}=4$ supermultiplet containing the  stress-energy tensor and the R-symmetry currents of the theory.

In a more realistic case, where conserved currents are used to define the detectors, the analysis goes via the same steps. The only 
technical complication in this case arises because the four-point
correlation functions involve a large number of different tensor structures (increasing with the spin of the conserved current), 
accompanied by {\it a priori} different functions of the conformal cross-ratios. For
our approach to work efficiently, these functions should admit a Mellin transform. We do not have an argument that this
property will hold in a generic CFT. Moreover, in the examples that we consider below, we find that the correlation functions can be
naturally split into the sum of two terms with one of them given by the  Born-level contribution and the second one encoding 
coupling-dependent contributions. It is only the second term that
admits a Mellin transform. The first term does not have the desired property and needs to be treated separately.\footnote{It is 
interesting to note that in conformal OPE for correlation functions
all conformal blocks admit a Mellin transform (see e.g. \cite{Costa:2012cb}). This suggests that our approach can be employed to 
compute the charge flow correlations in terms of invariant CFT
data by applying the procedure of this section to each conformal block individually. Since it involves integration of the correlation
 functions over a time at spatial infinity, the convergence of the
OPE expansion could be a potential subtlety. The question whether this actually happens and whether it makes sense to talk about 
the contribution of a given primary operator to the charge flow
operators will be considered elsewhere.}

\subsection{Definition and kinematics}

As we mentioned in the Introduction, we are interested in studying the Wightman functions entering \re{Wightman}. Their definition involves the notion of a flow operator,
or of a detector, i.e.,  an integrated local operator insertion at spatial infinity that measures the flow of quantum numbers in a given direction.
Conventional choices for such operator include the energy-momentum tensor and other conserved Noether currents such that the
detectors measure the flow of the energy/momentum and global charges, respectively.\footnote{In the more familiar case, the insertion of $\int d^3 \vec x \,T_{0\mu}\xi^\mu$ 
measures the charge of the state associated to a conformal Killing vector $\zeta^{\mu}$. Here we are
interested in less inclusive observables.}
In general, nothing prevents us from sending any operator to spatial infinity to define a detector. However, since this will not be associated with a conserved charge, its
physical interpretation as a  charge-flow measurement will be lacking.

The simplest flow operators are of the scalar type. They can be defined by analogy with the energy flow operators 
\cite{Sveshnikov:1995vi,Korchemsky:1997sy,Korchemsky:1999kt,BelKorSte01} as
follows
\begin{align}\label{detector1}
{\cal O}(n) = \int_0^{\infty} d t \lim_{r \to \infty} r^2\, \widetilde{O} (t, r \vec n)
\, ,
\end{align}
where the light-like vector $n=(1,\vec n)$ (with $\vec n^2=1$) defines the orientation of the detector in Minkowski space-time.
The large $r$ limit in \re{detector1} corresponds to the fact that we put the detector at spatial infinity. The $t$ integral runs
over the detector working time. In general, we expect radiation to arrive both to null and time-like infinity (if it is carried away by massless and massive particles,
respectively).

\begin{figure}[h!t]
\psfrag{pi}[cc][cc]{$\scriptstyle \pi$}\psfrag{mpi}[cc][cc]{$\scriptstyle -\pi$}
\psfrag{u}[cc][cc]{$\scriptstyle \theta$}\psfrag{v}[cc][cc]{$\scriptstyle \tau$}\psfrag{0}[cc][cc]{$\scriptstyle 0$}
\psfrag{I1}[cc][cc]{ $\scriptstyle i^+$}
\psfrag{I2}[cc][cc]{ $\scriptstyle{\cal I}^+$}
\psfrag{I3}[cc][cc]{ $\scriptstyle i^0$}
\psfrag{r}[cc][cc]{$\scriptstyle |\vec x|=r$}
\centerline{\includegraphics[width = 0.65 \textwidth]{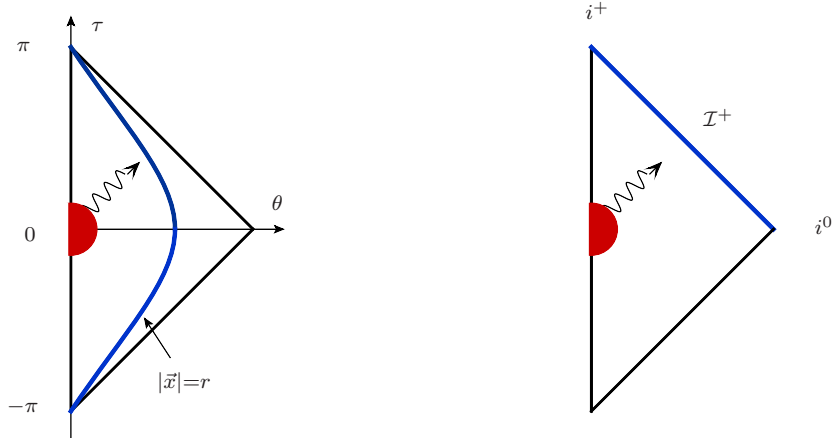}}
\caption{\small Penrose diagram of Minkowski space, $\tan({\tau \pm \theta \over 2})=t \pm |\vec x|$. a) We can consider a localized state (red) 
and first integrate over time and then take the large $r$ limit.
This procedure should work in theories of massive and massless particles. b) In CFTs the energy flux (as well as any other flux) from the state created by an
insertion of a local operator is carried away to the future null infinity, denoted as $\mathcal{I}^+$. In this figure, the symbols $i^+$ and $i^0$ 
stand for future time-like and spatial infinities, respectively.
It is convenient to work with momentum eigenstates which are not localized. For these reasons we adopt a procedure where we first send the 
detectors to the future null infinity ${\cal I}^+$ and then
integrate over the retarded (working) time.}
\label{fig-ee}
\end{figure}

In CFTs all radiation (energy flow) created by an insertion of a local operator goes to null
infinity.\footnote{This can be deduced from the form of the three-point function in CFT, e.g.,  $\vev{ O^{\dagger}(x) T_{\mu \nu}(x') O(0) }$
corresponding to $\widetilde{O} = T_{\mu\nu}$,  which does not possess any intrinsic mass scales.}
 Since all excitations carry energy, we expect that the same is true for all types of flow in CFTs and that they get
contribution only from null infinity. This explains why the limit in  \re{detector1} should be taken in such a way that the retarded time 
$t-r$ is kept fixed while the advanced time $t+r$ is sent to infinity (or
equivalently $\tau - \theta=\text{fixed}$ and $\tau+\theta \to\infty$, see Fig.\ \ref{fig-ee}).   The integral over $t$ in \re{detector1} becomes an 
integral over the whole range of the retarded time.  It is clear from \re{detector1}
that the flow operator does not depend on the lower limit of integration which can be chosen arbitrarily, so that the time translation invariance is not broken.%
\footnote{For localized states (states with finite support) the definition of the flow operator \re{detector1} is equivalent in CFT to the
one used in \cite{Hofman:2008ar}, $\mathcal O(n)=\lim_{r \to \infty} r^2 \int_{- \infty}^{\infty}   \widetilde{O}(t, r \vec n) $, where the time integral 
and the large distance limit are interchanged. In this
paper we are interested in states which are momentum eigenstates (which are not localized) so it is important to take the limit first as in \re{detector1}.}
Thus, by introducing the light-like four-vectors $n = (1, \vec n)$ and $\bar n = (1, - \vec n)$ we can rewrite \re{detector1} in a manifestly 
Lorentz covariant way\footnote{We work in Minkowski space
with signature $(+---)$.}
\begin{align}\label{detector10}
{\cal O}(n) &={1 \over 4} \int_{- \infty}^{\infty} d (xn) \lim_{(x \bar n) \to \infty} (x \bar n)^2 \widetilde{O} \left( {(x \bar n) \over 2} n + {(x n) \over 2} \bar n \right).
\end{align}
We can further generalize this expression to arbitrary null vectors $n$ and $\bar n$ with $(n \bar n) \neq 0$.
The flow operator is then defined as follows
\begin{align}\label{detector11}
{\cal O}(n) &= (n \bar n)  \int_{- \infty}^{\infty} d x_{-} \lim_{x_{+} \to \infty} x_+^2 \,\widetilde{O} \left( x_+ n + x_- \bar n \right)
\, ,
\end{align}
where we introduced the light-cone coordinates
\begin{align}
x_+ = {(x \bar n) \over (n \bar n)}
\, ,  \qquad
x_- = {(x n) \over (n \bar n)}
\, .
\end{align}
Choosing $n = (1, \vec n)$ and $\bar n = (1, - \vec n)$, the relation \re{detector11} reduces back to  \re{detector10}.

Equation \re{detector11} is the primary definition of the detector that we use below. We would like to stress that,
applying  \re{detector11}, we shall take the limit of the large advanced time $x_{+}$ first and integrate over $x_-$ afterwards.
The existence of the large $x_+ $ limit  in \re{detector11} is guaranteed by the properties of the conformal primary operator
$\widetilde{O}(x)$.\footnote{For a general conformal primary operator $\widetilde{O}$ with scaling dimension $\Delta$ and spin $s$, the definition \re{detector11} involves
a factor of $(x_+)^{\Delta - s}$, where the exponent is the twist of the operator.}
This can be seen by means of a global conformal transformation that maps
infinity to a finite distance \cite{Cornalba:2007fs, Hofman:2008ar, Zhiboedov:2013opa}.

Equipped with the above definitions, the scalar flow correlations are given by the following formula
\begin{align}\label{chargeflowcorr}
\la {\cal O}(n_1) \dots  {\cal O}(n_k) \ra_{q} ={\int d^4 x\, \e^{iqx}   \vev{0|  O^{\dagger}(x) {\cal O}(n_1) \dots {\cal O}(n_k) O(0)|0}
\over
\int d^4 x\, \e^{iqx}   \vev{0|  O^{\dagger} (x) O(0)|0}}
\, ,
\end{align}
where the subscript indicates the total four-momentum transfer $q^\mu$. In relation \re{chargeflowcorr}, the numerator is a particular
example of  $\sigma_w$ from \re{Wightman}, while the denominator is the total cross section $\sigma_{\rm tot}$ which corresponds to ${\cal D}[w] = 1$.

Defined in this way, the charge flow correlations \re{chargeflowcorr} do not depend on the
normalization of the operator $O$ that creates the state. Compared with the general definition \re{Wightman}, relation \re{chargeflowcorr} corresponds to a special 
choice of the flow operator $\mathcal D =  {\cal O}(n_1) \dots {\cal O}(n_k)$  with the light-like vectors $n_1,\dots,n_k$ specifying the orientations of the scalar detectors. 
Notice that the two-point function in the denominator of \re{chargeflowcorr} is non-zero only for time-like momenta with positive energy, $q^\mu q_\mu>0$ and $q^0>0$. 
It can be seen in many ways (see, e.g., Eq.~\re{SigmaTot} below) but
intuitively it is a kinematical constraint on the decay of a massive state with the total momentum $q^\mu$ into 
massless particles. The same is true about the numerator, even though it is not immediately
obvious in the form written above. Below we will always be interested in a time-like four-vector $q^\mu$ with a positive temporal component.

Replacing the flow operators in \re{chargeflowcorr} by their definition \re{detector11}, we find that $\la {\cal O}(n_1) \dots  {\cal O}(n_k) \ra_{q}$ 
can be obtained from the $(k+2)-$point correlation
function of scalar operators through the `detector procedure' which includes sending $k$ operators to infinity with subsequent integration 
over their position (retarded time) at the future null infinity.
One possible issue in \re{chargeflowcorr} is the order of different operations. For example, we can adopt a procedure where we first take the 
limit for one operator, integrate it over time, then
proceed to the second detector and so on. Alternatively, one can take all the limits first and then evaluate the time integrals. There is no 
{\it a priori} reason for these two procedures to be equivalent.
Indeed, we have found that the first route allows us to define charge flow correlations for any configuration of the detector vectors 
$n_1,\dots,n_k$, whereas the second one is ambiguous for $n_i = n_j$,
in which case the detector operators become light-like separated if the limit is taken first. We will be interested in the region $n_i \neq n_j$
 where both procedures yield the same answer.

Let us now examine the symmetries of the charge flow correlations. The charge flow operators are manifestly Lorentz invariant functions of $q^\mu$ and $n_i^\mu$,
\begin{align}\label{Lorentz}
\la {\cal O}(\Lambda n_1) \dots {\cal O}(\Lambda n_k) \ra_{\Lambda q} = \la {\cal O}( n_1) \dots {\cal O}(n_k) \ra_{q}
\, .
\end{align}
The second property follows from the dimensional analysis and conformal transformations  (dilatation symmetry) of the conformal primary operators,
 \begin{align}\label{additional}
\la {\cal O}(\lambda_1 n_1) \dots {\cal O}(\lambda_k n_k) \ra_{\lambda_q q}
=
\lambda_q^{-k} \prod_{i=1}^{k} \lambda_{i}^{-1} \la {\cal O}( n_1) \dots {\cal O}( n_k) \ra_{q}
\, .
\end{align}
A related property is the independence of \re{chargeflowcorr} on the auxiliary light-like vectors $\bar n_i$. Any dependence on the latter drops 
out upon taking the limit \re{detector11}. This is explicitly shown below.

The symmetries we have just discussed imply that the charge flow correlations are functions of the variables  $q^2 (n_i n_j)/ (2(q n_i) (q n_j))$ and
$(qn_i)(n_j n_l)/((qn_j)(n_i n_l))$, which are invariant under the transformations \re{Lorentz} and \re{additional}.
In particular, the one-point correlation is fixed up to an overall normalization constant $C$,
\begin{align}\label{oneandtwo1}
\la {\cal O}(n)\ra_{ q} = {C \over (q n)} \, ,
\end{align}
while the two-point correlation is a function of one variable,
\begin{align}\label{oneandtwo}
\la {\cal O}(n) {\cal O}(n')\ra_{q} &= {1 \over q^2 (nn')}  {{\cal F}(z) \over 4 \pi^2}
\, ,
\end{align}
with $z$ given by
\begin{align}\label{z}
z = {q^2 (n n') \over 2 (qn) (qn')}
\, .
\end{align}
We refer to the function $\mathcal{F} (z)$ as the event shape function. For a time-like $q^\mu$, the variable $z$ is restricted to the range $0 \leq z \leq 1$. In the rest frame of the source, $q^\mu = 
(q^0,\vec{0})$, it is related to the angle between the detectors, $z=(nn')/2=(1 - \cos\theta)/2$.

At the kinematical boundary $z = 0$, the detectors sit on top of
each other, $n=n'$. As was already mentioned,
we do not consider the degenerate case $n=n'$ in the present paper, even though it can be addressed with due care. In the vicinity of this point, for $z \ll 1$,
the detectors approach each other arbitrarily close in the angular direction on the sphere. This region was suggested to be governed by a version of the light-cone
operator product expansion \cite{Hofman:2008ar}.
The opposite case, $z=1$, can conveniently be visualized in the rest frame of the source, for
$\vec q = 0$, where it corresponds to detectors positioned on the north and south poles of the sphere, i.e., the so-called back-to-back kinematics. This case requires a
very careful consideration as well and we address it briefly below. The $z \to 1$ asymptotics of  the event shape function $\mathcal{F} (z)$ is governed by soft and collinear
emissions and has been studied at length in gauge theories \cite{Catani:1992ua,Korchemsky:1994is,Dokshitzer:1999sh}.

The relation \re{oneandtwo} is remarkably similar to the well-known expression for the two-point correlation function in a boundary CFT \cite{Liendo:2012hy}. Indeed, as was noticed in
\cite{Hofman:2008ar}, based on purely kinematical considerations, the charge flow correlations can be interpreted as expectation values of the product of primary operators in some
two-dimensional CFT over a nontrivial vacuum state $\ket{0}_q$. More precisely, each flow operator $\mathcal O(n_i)$ maps into a conformal primary field which carries the scaling
dimension $1$ and `lives' in the embedding space with the coordinates $n_i^\mu$ (subject to $n_i^2=0$ and identification of points $n_i^\mu  \to \lambda n_i^\mu$).\footnote{In the
gauge $n_i^0=1$, we can define the coordinates through the stereographic projection of the sphere $\vec n_i^2=1$.} The source defines the vacuum of two-dimensional CFT and
introduces a dependence on the `boundary vector' $q^\mu$ that breaks the two-dimensional conformal invariance.\footnote{Apart from kinematical similarities, there is no evidence for
the  existence of a dynamical local boundary CFT on the sphere at infinity that describes charge flow correlations of four-dimensional CFTs. Of course, if such a theory existed, it would be
a very exciting scenario \cite{Strominger:2013lka}.}

Our main focus will be on the double charge flow correlations \re{oneandtwo} in the interior of the region $0 \leq z \leq 1$.
As follows from the definition \re{chargeflowcorr}, the event shape function \re{oneandtwo} is determined by the four-point 
Wightman correlation function $\vev{O^\dagger (x)\widetilde O(x_2) \widetilde O(x_3) O (0)}$ involving the source and sink operators, $O$ and $O^\dagger$, as well as
two scalar operators $\widetilde{O}$ defining the flow operators \re{detector10}. All of them are conformal
primary fields with the scaling dimension $\Delta=2$. Conformal symmetry fixes the form of the correlation function in Euclidean space
to be \cite{Polyakov:1970xd}
\begin{align}\label{euclidfunction}
\vev{O^\dagger (x_1)\widetilde O(x_2) \widetilde O(x_3) O (x_4)} = {1 \over (2 \pi)^4} {\Phi(u,v) \over x_{12}^2 x_{13}^2 x_{24}^2 x_{34}^2}\,,
\end{align}
where $\Phi$ is a function of the two conformal cross-ratios,
\begin{align}
\label{ConformalCrossRatios}
u ={x_{12}^2 x_{34}^2 \over x_{13}^2 x_{24}^2}
\, , \qquad
v ={x_{23}^2 x_{14}^2 \over x_{13}^2 x_{24}^2}
\, .
\end{align}
Its explicit form depends on the dynamics of the underlying CFT.

\subsection{Analytic continuation}

In this subsection we explain  how to obtain non-time ordered, Wightman correlation functions from their Euclidean counterparts \re{euclidfunction}.
The relevant procedure of analytic continuation (Wick rotation) has been known for quite
some time. In the simplest case of scalar theories, we may follow the well-known Osterwalder-Schrader reconstruction theorem \cite{OstSch73}.\footnote{
The latter provides a
connection between the two functions by starting with a Euclidean path integral \cite{GlimJaffe}, that has a natural vector space structure, and recovering the Hilbert
space of states of the corresponding Wightman theory. Unfortunately, the procedure faces insurmountable
difficulties in the case of gauge theories, since the space of gauge equivalent connections does not have a natural vector space structure \cite{GeomConstrained}. However, conformal
gauge theories, not being theories of particles, are defined on a Hilbert space spanned by conformal primaries. Namely,
the Hilbert space of physical states of any such theory carries a unitary representation of the infinitely-sheeted universal covering group of the Minkowskian
conformal group \cite{LusMac75}.} This approach was extended by L\"uscher and Mack \cite{LusMac75} adapting it to the conformal field theory setup. The main conclusion of
Ref.~\cite{LusMac75} can be summarized in the following (simplified) statement: the correlation functions $G_E (x_1^4, \vec{x}_1; \dots ; x_n^4, \vec{x}_n) =
\vev{O(x_1^4, \vec{x}_1) \dots O(x_n^4, \vec{x}_n)}$ in Euclidean conformal field theory can be analytically continued to complex variables $x_k^4 = -\epsilon_k + i t_k$ 
in such a way that $G_E$ is holomorphic on the single-sheeted domain $0 < \epsilon_1 < \dots < \epsilon_n$, with $t_k$ kept arbitrary real. The formal limit $\epsilon_n \to 0$ 
defines a single-valued function of the time variables $t_1,\dots,t_n$, which coincides with the non-time ordered correlation function $\vev{O(t_1,\vec x_1)\dots O(t_n,\vec x_n)}$ 
in Minkowski space-time satisfying the Wightman axioms \cite{SteWig64}. Notice that unlike \cite{LusMac75}, we presently consider only a Poincar\'e patch on the cylinder 
$\mathbb{R} \times S^3$ where the CFT lives. 

To demonstrate the procedure let us consider, following \cite{LusMac75}, the example of a
two-point correlation function of conformal primary operators $O$ with scaling dimension $\Delta$. Conformal symmetry fixes the form of the Euclidean
correlation function up to an overall normalization (irrelevant for the argument that follows),
\begin{align}\label{LM-ex}
G_E (x_1^4, \vec{x}_1; x_2^4, \vec{x}_2) = \vev{O^{\dagger}(x_1)O(x_2)} = \frac{1}{[(x_1 - x_2)^{2}]^\Delta}
\, ,
\end{align}
where $x^2 = (x^4)^2 + \vec{x}^2$ in Euclidean signature. We immediately verify that for  the complex variables $x_k^4
= -\epsilon_k + i t_k$  the denominator in \re{LM-ex} cannot vanish in  the domain $0<\epsilon_1<\epsilon_2$ and $- \infty < t_k < \infty $.
Then the limit $\epsilon_k
\to 0$ with $\epsilon=\epsilon_2-\epsilon_1>0$ defines a tempered distribution,
\begin{align}\label{2pt}
G_W (t_1, \vec{x}_1; t_2, \vec{x}_2) = \frac{1}{\left[-(t_1 - t_2- i \epsilon)^2 + (\vec{x}_1 - \vec{x}_2)^2 \right]^{\Delta}}\,,
\end{align}
i.e., the Wightman two-point function.

The procedure can be repeated for a three-point correlation function but it
becomes technically more involved for correlators of four or more operators since they have complicated dependence on the coordinates.
For instance, the expression for the Euclidean four-point correlation function \re{euclidfunction} involves a function of two conformal cross-ratios $\Phi(u,v)$. The 
L\"uscher-Mack prescription defines its analytic continuation in the space of $u$ and $v$. The function that we get differs from the usual time-ordered one by 
the contributions of the discontinuities across the cuts in the
$x_{ij}^2-$planes.\footnote{One can question whether the L\"uscher-Mack procedure produces some Wightman functions but not the Wightman 
functions of the theory. We employed the conventional Schwinger-Keldysh technique \cite{Landau} to verify this by a direct computation in the explicit examples below.}
To see this, consider the difference between the time-ordered and Wightman
four-point functions and choose all distances but $x_{12}^2$ to be  space-like,
\begin{align}\label{diff}
\vev{ {\rm T} \big[  O^\dagger (x_1)\widetilde O(x_2) \widetilde O(x_3) O (x_4)  \big] }
&- \vev{  O^\dagger (x_1)\widetilde O(x_2) \widetilde O(x_3) O (x_4) }
\\[2mm] \nonumber
&
= \theta(x^0_{21}) \theta(x_{12}^2) \vev{ [O^\dagger (x_1), \widetilde O(x_2)] \widetilde O(x_3) O (x_4) }
\, .
\end{align}
Here we took into account that two operators, separated by a space-time interval, commute.
The time-ordered correlation function is obtained from its Euclidean analogue by a Wick rotation, $x_{ij}^2 \to - x_{ij}^2 + i \eps$,
but to reconstruct the Wightman function we have to examine in addition a large number of discontinuities. The situation becomes even more complicated if we
recall that the detector operators are integrated over their time coordinates.

As was observed by
Mack \cite{Mack:2009mi}, the above difficulties of the analytic continuation can be easily overcome by making use of the Mellin representation of the correlation
functions \cite{Symanzik:1972wj,Mansouri:1973zd,Dobrev:1975ru}. The Mellin representation is a natural language for the description of CFT correlators and came 
into a widespread use in recent years (see, e.g., \cite{Arutyunov:2000ku,Penedones:2010ue,Fitzpatrick:2011ia,Paulos:2011ie}).
For the connected part of the four-point Euclidean correlation function\footnote{\label{ConnectedFootnote}It is
defined as  $\la  O(x_1)  O(x_2)   O(x_3) O (x_4)  \ra_{c}  =\la O(x_1) O(x_2) O(x_3) O(x_4) \ra - \big[ \la O(x_1) O(x_2) \ra \la O(x_3) O(x_4) \ra +
{\rm perm} \big] $.}  it reads
\begin{align}\label{Mellin}
\vev{ O^\dagger (x_1)\widetilde O(x_2) \widetilde O(x_3) O (x_4) }_c = \int\prod_{1\le i<j\le 4} {d \delta_{ij} \over 2\pi i} (x_{ij}^2)^{\delta_{ij}} M(\delta_{ij})\,,
\end{align}
where the integration parameters satisfy the additional conditions $\sum_{j\neq k} \delta_{jk} = -\Delta$ with $\delta_{jk} = \delta_{kj}$ and $\Delta$ is the scaling dimension of
the operators. One of the advantages of the representation \re{Mellin} is that the power-like  short-distance singularities of the correlation function are translated into poles of the
Mellin amplitude $M(\delta_{ij})$. For our purposes the main convenience comes from the fact that the Mellin representation allows us to easily switch between Euclidean and
Lorentzian correlation functions.

Let us start with the Euclidean four-point correlation function \re{euclidfunction} of scalar operators with the scaling dimension $\Delta=2$. Then, the Mellin parameters in \re{Mellin}
can be chosen as $\delta_{12}=\delta_{34}=-1+j_1$, $\delta_{13}=\delta_{24}=-1+j_2$ and $\delta_{23}=\delta_{14}=-j_1-j_2$ leading to \cite{Mansouri:1973zd,Dobrev:1975ru}
\begin{align}\label{euclidmellin}
\Phi (u,v) = \int_{- \delta - i \infty}^{- \delta + i \infty} {d j_1 d j_2 \over (2 \pi i)^2}
M(j_1, j_2) \left({x_{12}^2 x_{34}^2\over x_{23}^2 x_{14}^2}\right)^{j_1} \left({x_{13}^2 x_{24}^2\over x_{23}^2 x_{14}^2}\right)^{j_2} \,.
\end{align}
Here the integration contours run along the imaginary axis with   $\Re j_{1,2} <0$ and  $\Re (j_{1}+j_{2})> -1$.
The Bose symmetry of the correlation function \re{euclidfunction} under the exchange of the operators at points $x_2$ and $x_3$ leads to the property $M(j_1, j_2) = M(j_2 , j_1)$.
Pulling out the product of Euler gamma functions that are usually included in the definition of the Mellin amplitude,\footnote{Their origin is explained in Appendix \ref {Dfunctions}
and stems from the Symanzik `star formula' \re{SymanzikForm}. Upon a Mellin transformation, each distance squared $x_{ij}^2$ is accompanied by an Euler gamma function.
For four-point correlation function this leads to a product of six gamma functions.}
we can write it in the form
\begin{align}\label{euclidmellinB1}
M(j_1, j_2)=\left[\Gamma(1-j_1) \Gamma(1 - j_2) \Gamma(j_1 + j_2)\right]^2
\widetilde M(j_1, j_2)\,,
\end{align}
where $\widetilde M(j_1, j_2)$ is the so-called amputated Mellin amplitude.

In the case of charge flow correlations built out of conserved currents, one can show that the disconnected part of the correlator does not contribute to the observables. In
the case of scalar flow operators at hand the disconnected part could lead to unphysical divergences, due to the presence of the unit operator
in the operator product expansion of the two detector operators. This is discussed in detail in \cite{paper2}.

Now, let us apply the L\"uscher-Mack analytic continuation to the Mellin integral \re{euclidmellin}.
As was demonstrated by Mack \cite{Mack:2009mi}, it amounts to replacing all Euclidean distances in the Mellin integrand by
\begin{align}
x_{ij}^2 \to - x_{i j}^2 + i \eps x^0_{i j}\theta(j-i)\,,
\end{align}
where the minus sign is due to our choice of the Minkowski signature $(+,-,-,-)$ and the $\theta-$function reflects the ordering of the operators
in  the Wightman function.\footnote{Time-ordered correlators can be similarly obtained by $x_{ij}^2 \to - x_{i j}^2 + i \eps$ .}
As was shown in  \cite{Mack:2009mi}, this procedure leads to correlation functions that satisfy the Wightman axioms \cite{SteWig64}.

\subsection{Integration over time and the final formula}
\label{LimitTime}

We are now ready to compute the charge flow observables \re{chargeflowcorr}. As explained in the beginning of this section,
the one-point correlation \re{oneandtwo} is fixed by symmetry. In the first nontrivial case of the two-detector correlation, we find from \re{detector11} and \re{chargeflowcorr}
\begin{align} \notag \label{OO}
\la {\cal O}({n})   {\cal O}({n}') \ra_{q}
&= \sigma_{\rm tot} ^{-1} \int d^4 x_1\, \e^{iqx_1} \int_{- \infty}^{\infty} d x_{2-} \int_{- \infty}^{\infty} d x_{3-}
\\
&\times
\lim_{x_{2+} \to \infty}  \lim_{x_{3+} \to \infty} (x_{2+} x_{3+})^2\vev{0|  O^{\dagger}(x_1) \widetilde{O}(x_2) \widetilde{O}(x_3)O(0)|0} \,,
\end{align}
where $x_2=x_{2+} n + x_{2-} \bar n$ and $x_3=x_{3+} n' + x_{3-} \bar n'$ define the positions of the detectors. The normalization factor $\sigma_{\rm tot}$
is given by the Fourier transform of the two-point Wightman correlator.
As mentioned in the Introduction, it describes the transition $O\to \text{everything}$. The correlation function
$\vev{0|  O^{\dagger} (x) O(0)|0}$ is completely fixed by conformal symmetry, Eq.~\re{2pt}, and its analytic continuation has been discussed in the previous subsection,
\begin{align}
\label{SigmaTot}
\sigma_{\rm tot} = \int {d^4 x\, \e^{iqx} \over (-x^2 + i\epsilon x_0)^2  } = 2\pi^3 \theta(q_0) \theta(q^2)\,.
\end{align}
Here we normalized the scalar operators so that their Euclidean two-point function is $1/(x^2)^2$.

To evaluate \re{OO}, we use the Mellin representation for the Wightman four-point correlation function,
\begin{align}\label{G4-W}
\vev{0|  O^{\dagger}(x_1) \widetilde{O}(x_2) \widetilde{O}(x_3)O(x_4)|0}_{c} = {1 \over (2 \pi)^4}
\int_{- \delta - i \infty}^{- \delta + i \infty} {d j_1 d j_2 \over (2 \pi i)^2} { M(j_1, j_2) \over  \hat x_{12}^2 \hat x_{13}^2 \hat x_{24}^2 \hat x_{34}^2 }  \left({\hat x_{12}^2 \hat x_{34}^2\over
\hat x_{23}^2 \hat x_{14}^2} \right)^{j_1} \left({\hat x_{13}^2 \hat x_{24}^2\over \hat x_{23}^2 \hat x_{14}^2}\right)^{j_2},
\end{align}
where $\hat x_{ij}^2 = -x_{ij}^2 + i\epsilon x_{ij}^0$ for $i<j$. The calculation of \re{OO} is greatly simplified by the assumption that the integration over 
the Mellin parameters can be postponed till the
very end, after the limit has been taken and all other integrations on the right-hand side of  \re{OO} have been done. We start by taking the limit. 
Replacing $x_2=x_{2+} n + x_{2-} \bar n$ and
$x_3=x_{3+} n' + x_{3-} \bar n'$ in \re{G4-W}, we find that for $x_{2+}, x_{3+} \to\infty$ the integrand on the right-hand side of \re{G4-W} scales as
\begin{align}\label{G4-integrand}
\frac14 (x_{2+} x_{3+})^{-2}
   M(j_1, j_2) \left[ 2 (\hat x_{12}n) (\hat x_{34}n')  \right]^{j_1-1}
  \left[ 2 (\hat x_{13}n) (\hat x_{24}n')  \right]^{j_2-1}  [\hat x_{14}^2 (nn') ]^{-j_1-j_2} \,,
\end{align}
where $(\hat x_{jk}n) \equiv (x_{j} n)- (x_kn) -i\epsilon$ for $j<k$ and $(\hat x_{jk}n')$ is defined in the same way.
Here we  tacitly assumed that $(n n') \neq 0$. The case $n = n'$, i.e., where the two detectors sit on top of each other, has to be treated separately.

Substituting \re{G4-W} and \re{G4-integrand} into \re{OO}, we integrate over the detectors times, $x_{2-}$ and $x_{3-}$,
and obtain an expression for the two-point correlation $\la {\cal O}(n)   {\cal O}(n') \ra_{q}$. It takes
the expected form \re{oneandtwo} with the {event shape function} given by (see appendix A for details)
\begin{align}\label{fourierrep}
{\cal F}(z) &=q^2 \int d^4 x \,{\rm e}^{i q x}  {   {\cal G}(\gamma)\over x^2-i\epsilon x^0} \,,
\end{align}
where $x\equiv x_{14}$ and the variable $z$ was defined in \re{z}. Here the integrand involves the function
\begin{align}\label{calG}
{\cal G}(\gamma)
=
- {1 \over 16 \pi^3}  \int_{- \delta - i \infty}^{- \delta + i \infty} {d j_1 d j_2 \over (2 \pi i)^2}
\left[{ \Gamma(1-j_1 - j_2)  \over \Gamma(1 - j_1) \Gamma(1 - j_2)}\right]^2 M(j_1 , j_2)  \gamma^{j_1 + j_2-1}\,,
\end{align}
which depends on the dimensionless ratio
\begin{align}\label{gamma}
\gamma = {2 ( (xn) - i \epsilon) ( (xn') - i \epsilon) \over  (x^2-i\epsilon x^0) (nn') }
\, .
\end{align}
Notice that the $(-i\epsilon)$-prescription in \re{fourierrep} and \re{gamma} follows unambiguously from the analytic properties
of the Wightman correlation function \re{G4-W}. In particular, it ensures that the function ${\cal F}(z)$ vanishes outside of
the physical region $0 \leq z \leq 1$.

Taking the Fourier transform in \re{fourierrep} we  finally obtain for $0<z<1$
\begin{align}\label{weightedcross}
{\cal F}(z)
= {1 \over 4} \int_{- \delta - i \infty}^{- \delta + i \infty} {d j_1 d j_2 \over (2 \pi i)^2} M(j_1, j_2)  K(j_1, j_2; z)\,,
\end{align}
where the $z-$dependence resides in the kernel
\begin{align}\label{K}
K(j_1, j_2; z)
= {2 \Gamma(1-j_1 - j_2) \over  \Gamma(j_1 + j_2)\left[ \Gamma(1 - j_1) \Gamma(1 - j_2)\right]^2 } \left({z \over 1 - z} \right)^{1 - j_1 - j_2}\,,
\end{align}
and $M(j_1, j_2)$ is the Mellin amplitude defining the Euclidean correlation function \re{euclidmellin}.

The following comments are in order. In the course of the derivation of \re{weightedcross} we defined several integrals by analytic continuation from their regions of validity.
In addition, we exchanged the order of different operations, e.g., the Fourier integral in \re{OO} with the Mellin integral. This could possibly lead to some subtleties and, indeed, 
as we show below, the relation  \re{weightedcross} does not reproduce correctly the contribution to ${\cal F}(z)$ localized at $z=1$.
We  explain in Section \ref{MellinSection} how it can be recovered and present an explicit example in Appendix \ref{SingularTermsAppendix}.

One may wonder whether the integral in \re{weightedcross} is convergent.  We recall that, compared to the correlation functions, here we have no 
{\it a priori} reason for \re{OO} to be finite. To address this issue it is convenient to rewrite \re{weightedcross} in terms
of the amputated Mellin amplitude $\widetilde M$ defined in \re{euclidmellinB1},
\begin{align}\label{weightedcrossB}
{\cal F}(z)
=
{1 \over 2 }
\int_{- \delta - i \infty}^{- \delta + i \infty} {d j_2 \over 2 \pi i}   {\pi \over \sin(\pi j_2)}  \left({z \over 1 - z} \right)^{1 - j_2}
\int_{- \delta - i \infty}^{- \delta + i \infty} {d j_1 \over 2 \pi i} \widetilde M(j_1, j_2 - j_1) \,,
\end{align}
where we shifted the integration variable as $j_2 \to j_2 - j_1$. The  usual assumption about the convergence of the Mellin amplitude $\widetilde M(j_1, j_2)$ is
that it does not grow exponentially fast for large $j_{1,2}$. Then, due to the damping multiplier $1/\sin(\pi j_2)$, the integral over
$j_2$ in \re{weightedcrossB} is convergent. This does not guarantee, however, the convergence of the $j_1-$integral.
The condition for the {event shape function} ${\cal F}(z)$ to be finite translates into the requirement for $\widetilde M(j_1, j_2 - j_1) $ to decrease sufficiently fast
at large $j_1$,
\begin{align}\label{irsafety}
\lim_{j_1 \to \infty }  \widetilde M(j_1, j_2 - j_1)  = o(1/ j_1)
\, .
\end{align}
We are not aware of any necessary and sufficient conditions for such a behavior in generic CFT. In practice, energy correlations are believed to be finite non-perturbatively.
It will be interesting to understand what property of the correlation functions involving stress tensors guarantees the IR finiteness of  the event shape function ${\cal F}(z)$ 
in terms of the dynamical CFT data (three-point functions and anomalous dimensions of operators).

\subsection{Double discontinuity}

Relation \re{weightedcross} establishes a correspondence between the two-detector correlation $\la {\cal O}({n}) {\cal O}({n}') \ra_{q}$ and the
four-point correlator of scalar operators in the CFT. Namely, the {event shape function} $\mathcal F(z)$ is given by the convolution of the Mellin amplitude
$M(j_1,j_2)$ defining the four-point correlator and the universal detector kernel $K(j_1,j_2;z)$ independent of the details of the CFT.

Let us turn the logic around and ask the following question: what properties of the correlation functions are probed by the {event shape function} $\mathcal F(z)$?
For this purpose,  in this subsection we derive a relation between the two functions which does not require going through the Mellin representation and
which is formulated directly in configuration space.

To begin with, let us examine the functions $\mathcal G(\gamma)$ and $\mathcal F(z)$ introduced in \re{calG} and \re{weightedcross}, respectively. They
are defined in two different representations and are related to each other through the Fourier transform \re{fourierrep}. Unlike $\mathcal F(z)$,
which is defined in the interval $0<z<1$, the function $\mathcal G(\gamma)$ depends on the (unconstrained) variable $\gamma$ given in \re{gamma}.
The analytic properties of $\mathcal G(\gamma)$ follow from the integral representation \re{calG}. In particular, examining the convergence properties of the Mellin
integral in \re{calG}, we find that $\mathcal G(\gamma)$ has a branch cut that runs along the negative axis $-\infty <\gamma\le 0$. We can define the discontinuity
of $\mathcal G(\gamma)$ across the cut as
\begin{align}
{\rm Disc}_\gamma \,\mathcal G(\gamma) \equiv { \mathcal G(\gamma+i0)- \mathcal G(\gamma-i0) \over 2i}\,,
\end{align}
{where the expression in the right-hand side is given by the difference of the functions evaluated at the two edges of the cut.} Making use of the identity
${\rm Disc}_\gamma\,  \gamma^{j_1+j_2} = \theta(-\gamma)|\gamma|^{j_1+j_2} \sin(\pi(j_1+j_2))$,
we find from \re{calG}
\begin{align}
{\rm Disc}_\gamma\, \mathcal G(\gamma) =  {\theta(-\gamma)\over 16 \pi^2 }   \int_{- \delta - i \infty}^{- \delta + i \infty} {d j_1 d j_2 \over (2 \pi i)^2}
{ \Gamma(1-j_1 - j_2) \, M(j_1 , j_2) \over \Gamma(j_1+j_2)[\Gamma(1 - j_1) \Gamma(1 - j_2)]^2}  (- \gamma)^{j_1 + j_2-1}  \, .
\end{align}
 We observe the striking similarity of this relation with \re{weightedcross} and \re{K}. Comparing the two integrals we conclude that
 \begin{align}\label{lastFourier}
{{\cal F}(z) \over  8 \pi^2} &= {\rm Disc}_\gamma \,\mathcal G(\gamma)\bigg|_{\gamma =  (z-1)/z}\,.
\end{align}
Thus, the role of the Fourier integral in \re{fourierrep} is to take the discontinuity of $\mathcal G(\gamma)$ across the cut and then to set {$\gamma = (z-1)/z$}.

As a next step, we have to find a relation between $\mathcal G(\gamma)$ and the four-point correlation function \re{euclidfunction}, or equivalently,
the function $\Phi(u,v)$ of the two conformal cross-ratios. In Euclidean space, $u$ and $v$ take positive values and the Mellin amplitude is given by
(see Eq.~\re{euclidmellin})
\begin{align}\label{invmellin}
M(j_1 , j_2) = \int_0^\infty {d u d v\over v^3} \Big( {u \over v} \Big)^{-1-j_1} \Big( {1 \over v} \Big)^{-1-j_2} \Phi(u,v)\,.
\end{align}
We may try to substitute this relation into \re{calG} and  \re{weightedcross} but it is easy to check that the resulting Mellin integrals diverge.
This is not surprising since otherwise it would imply that the intrinsically Lorentzian {event shape function} $\mathcal F(z)$ can be
expressed in terms of a Euclidean quantity.

It is therefore more natural to relate the Mellin amplitude to the function $\Phi(u,v)$ in Minkowski space-time where $u$ and $v$ can take arbitrary real values.
It is convenient to parametrize the possible values of the cross-ratios as
\begin{align}\label{crossratios}
u = {w \bar w \over (1-w)(1-\bar w)}\,,\qquad v = {1 \over (1-w)(1-\bar w)}\,,
\end{align}
so that the corresponding function $\Phi (w, \bar w)\equiv \Phi\left(u(w,\bar w),v(w,\bar w)\right)$ has the Mellin representation
\begin{align}\label{euclidmellinB}
\Phi (w,\bar w) = \int_{- \delta - i \infty}^{- \delta + i \infty} {d j_1 d j_2 \over (2 \pi i)^2} M(j_1, j_2) \left(w \bar w \right)^{j_1} \left((1- w)(1 - \bar w)\right)^{j_2}\,.
\end{align}
On general grounds we expect that $\Phi (w,\bar w)$ should have cuts in the $w-$ and $\bar w-$planes that start at the branch points at $w = 0,1$ and
$\bar w=0,1$. We can choose the cuts to be $(- \infty, 0]$ and $[1, \infty)$. The principal sheet is the one on which the Euclidean correlation function
is defined. As before, we may use \re{euclidmellinB} to compute the discontinuity of $\Phi (w,\bar w)$ across the cuts. For our purposes we will need
the  Lorentzian double discontinuity ${\rm Disc}_{w}{\rm Disc}_{\bar w} \Phi(w, \bar w)$ across two different cuts, $-\infty < w\le 0$ and $1\le \bar w<  \infty$
\footnote{Discontinuities of the Lorentzian correlation functions in CFT have been studied in the context of locality \cite{Gary:2009ae} and the eikonal approximation \cite{Cornalba:2006xm} in AdS/CFT.}
\begin{align}\notag\label{disc}
{\rm Disc}_{w}{\rm Disc}_{\bar w} \Phi(w, \bar w) &= - \theta(-w)\theta(\bar w-1)\int_{- \delta - i \infty}^{- \delta + i \infty} {d j_1 d j_2 \over (2 \pi i)^2} M(j_1, j_2)
\\
& \times
 \left(-w \bar w \right)^{j_1} \left((1- w)( \bar w-1)\right)^{j_2}\sin(\pi j_1)\sin(\pi j_2)\,.
\end{align}
The inverse relation looks as
\begin{align}\label{M-imp}
M(j_1, j_2) =  \int_{- \infty}^{0} d w  \int_{1}^{\infty} d \bar w   {(w - \bar w) \,
{\rm Disc}_{w}{\rm Disc}_{\bar w} \Phi(w, \bar w)  \over  \left(-w \bar w \right)^{1+j_1} \left((1- w)( \bar w-1)\right)^{1+j_2}\sin(\pi j_1)\sin(\pi j_2)}  \,.
\end{align}
Compared to \re{invmellin}, this integrand contains the additional factor $1/(\sin(\pi j_1)\sin(\pi j_2))$ which improves the convergence of the Mellin
integrals at infinity. Indeed, substituting \re{M-imp} into \re{calG} and performing the integration with respect to the Mellin parameters, we find
 \begin{align}\label{doubledisc}
{\cal G}(\gamma) &= -{1 \over 16 \pi^5}  \int_{- \infty}^{0} d w  \int_{1}^{\infty} d \bar w   {w - \bar w \over w \bar w (1-w)(1 - \bar w)}
{\rm Disc}_{w} {\rm Disc}_{\bar w}  \Phi(w, \bar w) \\ \nonumber
&\times \left( {1 \over \sqrt{(-1 + w + \bar w +\gamma)^2 - 4 w \bar w \gamma}} +{1 \over (1-w)(1-\bar w) - \gamma} +{1 \over w \bar w - \gamma} + {1 \over \gamma} \right),
\end{align}
where the discontinuities are taken along the left branch of the cut in the $w-$plane and along the right one in the $\bar w-$plane.

Finally, we combine Eqs.\ \re{lastFourier} and \re{doubledisc} together to conclude that the {event shape function} $\mathcal F(z)$  is related to the double discontinuity of the
four-point correlation function. We observe that the $\gamma$ dependence resides only in the second line of \re{doubledisc} and, as a consequence, each of the four
terms in the paranetheses in \re{doubledisc} contributes to the discontinuity in \re{lastFourier}. In particular, the last term there, $1/\gamma$, gives rise
to a contribution to $\mathcal F(z)$ proportional to $\delta(1-z)$. This suggests that relation \re{lastFourier} can be used to correctly reproduce the
$\delta(1-z)$ terms in the {event shape function} $\mathcal F(z)$.  Below we present an example illustrating this point.

\section{From correlation functions to amplitudes}
\label{CorrelatorToAmplitude}

In the previous section we analyzed the charge correlations in a generic interacting four-di\-men\-sional CFT and our discussion did not rely on  whether 
an S-matrix formulation exists or not. Typically, the latter is not available since the theory remains interacting at any distances. If we insist on defining a scattering matrix to make
contact with QCD-like theories, the way out of this predicament is to deform the original theory away from its
critical point. There are multiple ways to do it, either by deforming the theory with a relevant operator that triggers the flow to the gapped phase or by deviating from the
four-dimensionality of space-time. Along this route, the theory becomes free in the IR and thus the flow operators acting at infinity can be regarded as acting on the Hilbert space
of free massless particles. The definition of scattering amplitudes becomes meaningful in this setup and it is safe to talk about weighted cross sections \re{event-shapes}. The
aforementioned IR divergences show up as singularities in the scattering amplitudes, either as logarithms in the mass or as poles in the parameter of dimensional 
regularization. However, the
fact that we are dealing with IR safe observables implies that all singularities cancel on the right-hand side of \re{event-shapes} and a finite limit exists when 
we send the deformation parameter to zero. There is one subtlety though that we should address: does the finite result that we get in this way coincide with the computation 
in the original theory using correlation functions \re{chargeflowcorr}? It is clear that, at least perturbatively, this is the case. Non-perturbatively we do not have any 
arguments in favor of or against this scenario and it will be interesting to explore it in detail.

Having in mind perturbative computations around the free field theory fixed point, it suffices to explain the relation between the integrated correlation functions \re{Wightman} and
weighted cross sections \re{event-shapes} in the simplest case of a massless scalar field $\phi$. The derivation presented below is not new,
see, e.g., Ref.~\cite{Bauer:2008dt}. We also introduce a bilinear local operator $\widetilde{O} (x) = \phi^2 (x)$ of dimension two that serves as a toy model for the detector discussed in the previous section.
Since we are dealing with a free theory, we can use the mode decomposition for the free scalar field $\phi$ in terms of creation and annihilation operators,
\begin{align}
\label{creationandannih}
\phi(x) &= \int { d^3 \vec k \over 2 k^0 (2 \pi)^3 }
\big[
a^{\dagger} (k)  {\rm e}^{i k^0 t - i \vec{k} \vec{x}} + a (k)  e^{- i k^0 t + i \vec{k} \vec{x}}
\,
\big]
\, .
\end{align}
Here the on-shell condition $k^0 = |\vec{k}|$ is implied and the creation/annihilation operators obey the commutation relation  $[a (k) , a^{\dagger} (k') ] = (2 \pi)^3
2k^0 \delta^{(3)} (\vec{k}-\vec{k}')$. We substitute the expansion \re{creationandannih} into the definition \re{detector11} of the flow operator,
\begin{align}\label{limitcreat}
{\cal O}(n) &= \int_{- \infty}^{\infty} d (x n)  \lim_{r \to \infty} r^2  \phi^2 (x n + r , r \vec n)
\, ,
\end{align}
and take the large $r$ limit first. This produces a highly oscillating integrand and we can evaluate the two angular integrals emerging from the momentum integration
measure $d^3 \vec{k} = \vec{k}^2 d |\vec{k}| d \Omega_{\vec{k}}$ by a stationary phase
\begin{align}\label{limitangle}
\lim_{r \to \infty} r \int d^2 \Omega_{\vec{k}} \, \e^{\pm i k r (1 - (\vec{n} \vec{k})/|\vec{k}|)}
=
\pm {2 \pi i \over |\vec{k}|} \int d^2 \Omega_{\vec{k}} \, \delta^{(2)} (\Omega_{\vec n} - \Omega_{\vec{k}})
\, .
\end{align}
Finally, the retarded $(x n)-$time integral produces the energy conserving delta function in the large $r$ limit. We can use it to integrate over $|\vec k '|$. The result
of these operations yields the final expression for the flow operator
\begin{align}\label{limitcreat2}
{\cal O}(n) &= \int { d |\vec{k}| \over 2 (2 \pi)^3 } a^{\dagger}  (k) a (k)
=
\int {d^3 \vec k \over 2 k^0 (2 \pi)^3} a^{\dagger} (k) a (k) \left({1 \over k^0} \delta^{(2)} (\Omega_{\vec n} - \Omega_{\vec k})  \right) .
\end{align}

Making use of \re{limitcreat2}, we find that the action of the flow operator ${\cal O}(n)$ on a multiparticle state $\ket{X}$ is diagonal,
with the corresponding eigenvalue being the weight factor $w_{{\cal O}} (X)$ (see Eq.\ \re{event-shapes}),
\begin{align}
\label{OdiagW}
{\cal O}(n) \ket{X}
=
w_{{\cal O}(n)} (X) \ket{X} = \sum_{i}  {1 \over k^0_i} \delta^{(2)} (\Omega_{\vec n} - \Omega_{\vec k})   | X \ra\,,
\end{align}
where the sum runs over all particles populating $\ket{X}$. Inserting the complete set of on-shell states, $\sum_X \ket{X} \bra{X} = 1$, in Eq.\ \re{chargeflowcorr} and naively using relation \re{OdiagW} multiple times
yields the equality of the scalar flow correlation and the weighted cross section,
\begin{align}\label{weightcrossAmp1}
\la {\cal O}(n_1) \dots {\cal O}(n_k) \ra_{q}
=
{1 \over \sigma_{\rm tot}}  \sum_{X}  (2 \pi)^4 \delta^{(4)}(q - k_{X}) w_{ {\cal O}(n_1)}(X) \dots w_{ {\cal O}(n_k)}(X)|\la X |O(0)|0 \ra |^2
\, .
\end{align}
However, taken at its face value, this identity is not correct beyond the one-point correlation.
Given our definition of the left-hand side  in Eq.\ \re{chargeflowcorr}, for the equality \re{weightcrossAmp1} to be valid the correlation 
is subject to certain conditions that we now address.

First, a careful inspection of the right-hand side of Eq.\ \re{weightcrossAmp1} shows that already in the free theory the scalar 
flow operators \re{limitcreat2} cease to commute. On the correlator
side, this complication translates into contributions from graphs where the flow operators are contracted by at least one Wick pairing, the so-called detector cross-talk. Thus, to
make the above identification \re{weightcrossAmp1}  work, one needs to eliminate the offending diagrams. This procedure is addressed at length in the companion
paper \cite{paper2}. 

Another issue that may arise is that
the correlations are ill defined because they are not associated with any conserved charge. Indeed, we found that the 
scalar correlations can diverge starting from two detector insertions.  Below
we provide an explicit illustration in $\mathcal{N}=4$ SYM where we overcome both  difficulties  by choosing very specific 
$SO(6)$ quantum numbers of  the scalar half-BPS bilinear operators
representing the sources and the flow operators. With this choice, the scalar detector correlations turn out to be related to the 
energy flow correlations  by superconformal symmetry. The latter
are free from the aforementioned complications since these effects are no longer present for flow operators built from the stress tensor.

\section{Illustration: ${\cal N}=4$ SYM}
\label{illustration}

In this section, we use $\mathcal N=4$ SYM  with the gauge group $SU(N_c)$ to illustrate the general formalism for computing charge 
flow correlations from Euclidean correlators as described above.
We recall that in a phenomenologically interesting setup the source is a gauge-invariant conserved current, e.g., an electromagnetic 
current. However, to avoid unnecessary complications due
to the Lorentz spin of the currents, we will deal with a scalar operator whose scaling dimension does not receive quantum corrections.   
Within the framework of $\mathcal{N}=4$ SYM theory
a natural candidate for such an object is the bilinear scalar half-BPS operator
\begin{align}\label{state}
O(x) = \frac{1}{\sqrt{c}} Y^{I} Y^{J} O_{\bf 20'}^{IJ} = \frac{1}{\sqrt{c}} Y^{I} Y^{J} {\rm tr}[\Phi^{I} (x) \Phi^{J} (x)] \, ,
\end{align}
where the six real scalar fields $\Phi^I$ (with $I=1,\dots,6$) form a vector of $SO(6)$. They are contracted with an auxiliary six-dimensional complex
null vector $Y^I$ in order to project the bilinear ${\rm tr}[\Phi^{I} (x) \Phi^{J} (x)]$ onto the irreducible representation ${\bf 20'}$ of the R-symmetry 
group $SU(4) \sim SO(6)$. We can choose without loss of generality
\begin{align}\label{4.2}
Y=(1,0,1,0,i,i)\,,
\end{align}
with any other choice being its $SO(6)$ rotation. Notice that we have included the factor
\begin{align}
c = (N_c^2 - 1)/(2 \pi^4)
\end{align}
into the definition \re{state} in order to have a convenient normalization for the Euclidean two-point function, $\vev{O^\dagger(x)O(0)}=1/(x^2)^2$.

The operator \re{state} belongs to the same $\mathcal N=4$ supermultiplet as the R-current  $J_\mu$ and the energy-momentum tensor 
$T_{\mu\nu}$ and hence its scaling dimension $\Delta=2$ is
protected from quantum corrections. As a consequence, {the four-point} correlation function of such scalar operators, discussed in this 
paper, is related by superconformal transformations to the
physically interesting {correlation functions} {involving conserved currents}. {Therefore, we may anticipate that also the corresponding 
flow operator correlations are related to one another, which is
indeed the case \cite{paper2}}.

To calculate the double scalar flow correlation, we consider the four-point function of the form $\vev{O^\dagger(x_1)
\widetilde O(x_2) \widetilde O(x_3) O(x_4)}$, where the operators at points $x_1$ and $x_4$ describe the source and its conjugate (the sink), respectively, and
$\widetilde O$ denotes the flow operator. The reality of the event shape function $\mathcal{F} (z)$ implies that the flow operators have to be Hermitian. For the case at hand it means
that the $SO(6)$ projectors have to be defined by real, symmetric and traceless matrices $S^{IJ}$,\footnote{Here the factor of $2$ in the definition of the operator $\widetilde O (x_2)$
was introduced to compensate a similar factor in the normalization of the $SU(N_c)$ generators ${\rm tr}(T^a T^b) = {1 \over 2} \delta^{a b}$.}
\begin{align}\label{detectors}
\widetilde O (x_2) = 2 S^{IJ} O_{\bf 20'}^{IJ}  = 2 S^{IJ} {\rm tr}[\Phi^{I} (x_2) \Phi^{J} (x_2)]
\, .
\end{align}
The second operator $\widetilde O (x_3)$ is given by the same expression with the matrix $S$ being replaced by another one $S'$ with the same properties.
It is convenient to choose $S$ and $S^\prime$ in the following form 
\begin{align}\label{detectorsS}
S = {\rm diag} (1, - 1, 0, 0, 0 ,0) \, , \qquad S' = {\rm diag} (0, 0, 1, -1, 0 ,0)\,,
\end{align}
so that $S S'=0$. This choice has an invariant interpretation in the sense of the R-symmetry group. Namely, for general $S$ and $S'$ the operator product expansion of
the two operators $\widetilde O(x_2) \widetilde O(x_3)$ results into six  $SU(4)$ channels corresponding to the irreducible representations  that appear in  the
tensor product ${\bf 20'} \times {\bf 20'} =  {\bf 1} +  {\bf 15} +  {\bf 20'} +  {\bf 84} +  {\bf 105} +  {\bf 175}$. However, for the particular choices of the auxiliary variables $Y$ in
\p{4.2} and $S, S'$ in \re{detectorsS}, only the representation  ${\bf 105}$ survives (for more details, see Appendix~D in Ref.~\cite{paper2}).

One of the reasons why we selected the contribution of the ${\bf 105}$ representation to the double scalar flow correlations is that, as we prove in 
Ref.\  \cite{paper3} using superconformal Ward identities, it is related to the physically interesting energy flow correlations in a remarkably simple way,
\begin{align}\label{relation}
\vev{ {\cal E} (n) {\cal E} (n') } = {4 (q^2)^2 \over (n n')^2} \vev{ {\cal O} (n) {\cal O} (n') }_{\bf 105}
\, .
\end{align}
Here the energy flow operator ${\cal E} (n)$ is determined by the same expression as before \re{detector11} with the only difference that the
scalar operator gets replaced by a particular component of the energy-momentum tensor, i.e., $\widetilde{O}(x)\ \to\  {\bar{n}^\mu \bar{n}^\nu} T_{\mu\nu} (x)/(n \bar{n})^2$.

We would like to emphasize that Eq.~\p{relation} does not rely on the specific dynamics of $\cN=4$ SYM, but it is a (rather non-trivial) 
corollary of  ${\cal N}=4$ superconformal symmetry. It allows us to use available results  for the correlation functions of half-BPS
operators in order to probe the energy flow correlations in ${\cal N}=4$ SYM  at weak and strong coupling. Another interesting 
consequence of Eq.\ \re{relation} is that $\vev{ {\cal O} (n) {\cal O} (n') }_{\bf 105}$ inherits the positivity property of the energy flow correlations. 
We shall return to this point in Sect.~\ref{sect-con}.

\subsection{Double flow correlation at weak coupling}

To start with, we recall that due to the half-BPS nature of the source \re{state}, the two-point Wightman correlation function 
$\vev{O^\dagger(x)O(0)}$ is given exactly by  $\vev{O^\dagger(x)O(0)}=1/(-x^2+i\epsilon x^0)^2$, Eq.\ \re{SigmaTot}, to all 
orders in perturbation theory.\footnote{See, e.g., \cite{Howe:1996rb,D'Hoker:1998tz,Howe:1998zi}
for non-renormalization theorems and Refs.\ \cite{Lee:1998bxa} and \cite{Penati:1999ba} for explicit one- and two-loop perturbative tests, respectively.}
This implies that the total cross section coincides with its Born level approximation, or equivalently only the tree-level  amplitude for the transition $O \to  \text{two scalars}$
contributes  to $\sigma_{\rm tot}$, with the rest cancelling exactly between real emissions and virtual corrections,
\begin{align}
\sigma_{\rm tot} = \int d^4 x\, \e^{iqx}   \vev{0|  O^{\dagger}(x) O(0)|0}  = (2 \pi)^2 \int d^4 k \, \delta_{+}(k^2) \delta_{+}((q-k)^2) = 2 \pi^3
\, .
\end{align}
Here the second relation is expressed in terms of a momentum integral of the Cutkosky cut involving two Wightman functions in  momentum space.

\subsubsection{Mellin transform}
\label{MellinSection}

Let us now examine the main object defining the double flow observable, the four-point Wightman correlation function of the half-BPS operators $O_{\bf 20'}^{IJ}$.
Its Euclidean counterpart was studied in great detail in $\mathcal{N}=4$ 
SYM at weak coupling \cite{Eden:2000bk, Heslop:2002hp,Dolan:2001tt,Eden:2012tu}, so we can adapt these results to our present consideration.

As was explained at the beginning of this section, to ensure the reality condition for the charge flow correlations, we consider the four-point
function $\vev{O^\dagger(x_1) \widetilde O(x_2) \widetilde O(x_3) O(x_4)}$ of the operators \re{state} and \re{detectors} involving different $SO(6)$ projections of the half-BPS
operator $O_{\bf 20'}^{IJ}$. Taking into account \p{4.2} and \re{detectors}, we find
\begin{align}\label{weakcoupl}
&
\vev{ O^{\dagger}(x_1) \widetilde O(x_2) \widetilde O(x_3) O(x_4) }
=
{N^2_c -1 \over 8 (2 \pi)^4} \left( {1 \over x_{12}^4 x_{34}^4} + {1 \over x_{13}^4 x_{24}^4 } \right)\\
&\qquad\qquad
+
{1 \over (2 \pi)^4 }
{1 \over x_{12}^2 x_{24}^2 x_{13}^2 x_{34}^2}
\left\{
\frac{1}{2} + \int_{- \delta - i \infty}^{- \delta + i \infty} {d j_1 d j_2 \over (2 \pi i)^2}
M(j_1,j_2; a)  \Big({x_{12}^2 x_{34}^2\over x_{23}^2 x_{14}^2}\Big)^{j_1} \Big({x_{13}^2 x_{24}^2\over x_{23}^2 x_{14}^2}\Big)^{j_2}
\right\}
\, , \nonumber
\end{align}
where the first and the second lines describe the disconnected and connected contributions, respectively (see footnote \ref{ConnectedFootnote}).
In addition, in the second line we separated the Born-level contribution from quantum-loop effects and made use of the Mellin representation for
the latter. The Mellin amplitude $M (j_1, j_2; a)$ depends on the 't Hooft coupling $a=g^2_{\scriptscriptstyle \rm YM} N_c/(4\pi^2)$ and starts at order $O(a)$
in perturbation theory \cite{Usyukina:1992jd},
\begin{align}
\label{OneLoopMellin}
M(j_1, j_2; a)&= -{a \over 4} \left[\Gamma(-j_1) \Gamma(- j_2) \Gamma(1+j_1 + j_2)\right]^2 + O (a^2)
\, .
\end{align}
For the amputated Mellin amplitude, we have from \re{euclidmellinB1}
\begin{align}\label{OneLoopMellinAmp}
\widetilde M(j_1, j_2; a) =  - a {(j_1+j_2)^2 \over 4 j_1^2 j_2^2 } + O (a^2)\,.
\end{align}
The integration contour in the Mellin integral in \re{weakcoupl} runs
along the imaginary axis  to the left of the origin and separates  the poles of the Euler gamma functions in \re{OneLoopMellin} with positive and negative signs in front
of the integration variables, i.e., $\Gamma (\ldots + j)$ and $\Gamma ( \ldots - j)$, respectively. Presently, the perturbative expansion of \re{weakcoupl} is known explicitly 
in terms of functions with known Mellin representation up to two loops \cite{GonzalezRey:1998tk,Eden:2000bk,Eden:2000mv,Bianchi:2000hn}, and up to six loops in 
terms of basis integrals \cite{Eden:2012tu},  which were successfully computed at three-loop order \cite{Drummond:2013nda}.

Let us now substitute \re{weakcoupl} into \re{OO} and compute the corresponding {event shape function} \re{oneandtwo}. 
At weak coupling, it admits an expansion in powers of $a$,
\begin{align}
\mathcal{F} (z;a) = \mathcal{F}^{(0)} (z) + a \,\mathcal{F}^{(1)} (z) + \dots
\, .
\end{align}
To obtain $\mathcal{F}^{(0)} (z)$ it suffices to retain only the terms independent of the coupling constant  in \re{weakcoupl}. 
We find in this way that the first line in Eq.\ \re{weakcoupl} does not
contribute to \re{OO}, since after sending the operators $\widetilde O(x_2)$ and $\widetilde O(x_3)$ (i.e., the two detectors) 
to spatial infinity, the integrals over the detector time take the
form $\int_{-\infty}^\infty d x_{2-} / (x_{2-} - i \eps)^2$ and hence vanish. For the Born level term in the second line of \re{weakcoupl}, 
the integral over the detector time involves a single pole and, therefore, yields a nontrivial result,
\begin{align}
\label{BornF}
\mathcal{F}^{(0)} (z)
=
- {q^2 (nn') \over 8 (2 \pi)^3} \int d^4 x {{\rm e}^{i q x} \over ((xn) - i 0) ((xn') - i 0)}
=
{1 \over 4} \delta(1-z)
\, .
\end{align}
At one loop, we apply the general strategy developed in Section \ref{LimitTime} and compute the {event shape function} $\mathcal{F}^{(1)} (z)$
using \re{weightedcross} and \re{K}. Taking into account \re{OneLoopMellinAmp}, we find
\begin{align}\label{weakcouplCHF}
&
\mathcal{F}^{(1)} (z)
=
-  {\pi \over 8}  \int_{- \delta - i \infty}^{- \delta + i \infty} {d j_1 d j_2 \over (2 \pi i)^2}
{ (j_1+j_2)^2\over j_1^2 j_2^2 \sin(\pi(j_1+j_2))} \left( {z \over 1 - z} \right)^{1 - j_1 - j_2}
= - {z \over 4} {\ln (1- z) \over 1 - z}
\, .
\end{align}
Notice that, as we pointed out in Section \ref{LimitTime},  the $O(a)$ correction in \re{weakcouplCHF} is valid only away from the kinematical boundaries, i.e., for
$0 < z < 1$. Indeed, a more careful analysis of the $z=1$ boundary (see Appendix \ref{SingularTermsAppendix} for details) leads to the following result 
\begin{align}
\label{Z1Subs}
\mathcal{F}^{(1)} (z)=
-{z \over 4} \left[ {\ln (1- z) \over 1 - z} \right]_{+} - {\zeta_2 \over 4} \, \delta (1 - z)\,,
\end{align}
where $[\dots]_+$ stands for the conventional definition of the plus-distribution, see Eq.\ \re{PlusPrescrip}. The two expressions
\re{weakcouplCHF} and \re{Z1Subs} differ by a contact term proportional to $\delta(1-z)$. Once such  terms are  taken into account,
the {event shape function} \re{Z1Subs} becomes integrable at $z=1$. This is consistent with the physical picture that the energy flow
correlations are finite when integrated over any finite region on the sphere.  Equivalently, we can obtain \re{Z1Subs} using \re{lastFourier}.  This is what we come to next.

\subsubsection{Double discontinuity}

As we already discussed at length in the previous section, there is yet another efficient way to obtain the result for the {event shape function} which immediately
yields contributions from the tempered distribution at $z = 1$. It relies on using Eq.\ \re{doubledisc}.

Substituting \re{OneLoopMellin} into \re{euclidmellinB}, we find that the one-loop conformal invariant function $\Phi (w, \bar w) = a\, \Phi^{(1)} (w, \bar w) + \dots$, rewritten in terms of the
variables $w$ and $\bar{w}$ defined in Eq.\ \re{crossratios}, takes the  form \cite{Usyukina:1992jd}
\begin{align}
\Phi^{(1)} (w, \bar{w})
= - 
\frac{1}{4(w - \bar{w}) }
\left[
2 {\rm Li}_{2} \left(w \right)
-
2 {\rm Li}_{2} \left(\bar{w} \right)
+
\ln (w \bar{w}) \ln \left( \frac{1 - w}{1 - \bar w} \right)
\right]
\, ,
\end{align}
with ${\rm Li}_2(x)$ being the Euler dilogarithm.  Its double discontinuity across the cuts $-\infty < w \le 0$ and $1\le \bar w<\infty$ reads
\begin{align}\label{Phi1}
{\rm Disc}_{w} {\rm Disc}_{\bar w}  \Phi^{(1)} (w, \bar w) =- {\pi^2 \over 4} {1 \over w - \bar w}
\, .
\end{align}
Substituting this relation into \re{doubledisc}, we can easily obtain the corresponding function ${\cal G}(\gamma)$ in the
coordinate representation,
\begin{align}
\label{calGoneloop}
{\cal G}(\gamma) = - {a \over 4 (2 \pi)^3 } {{\rm Li}_{2}({\gamma -1 \over \gamma}) - \zeta_2 \over \gamma} + O (a^2)
\, .
\end{align}
According to \re{lastFourier}, the {event shape function} can be computed as the discontinuity of ${\cal G}(\gamma)$ at $\gamma=(z-1)/z$. This discontinuity is easily
evaluated by replacing the dilogarithm by its integral representation (see Eq.~\re{GenFuncPoly} below). In this way we reproduce the expected result \re{Z1Subs},
including the terms localized at $z=1$. This should be compared with the conventional computation of similar observables using amplitudes,
which is quite intricate and requires different sorts of regularization \cite{Basham:1978zq}. The simplicity of the calculation of the scalar flow correlations
is one of the advantages of the current approach, provided the Euclidean correlation functions are known, which is currently the case.

\subsubsection{Weighted cross section}

Let us finally turn to the computation of the scalar flow correlations starting from $\mathcal{N}=4$ SYM amplitudes.  As explained in Section
\ref{CorrelatorToAmplitude}, the flow operators can be expressed in terms of the creation and annihilation operators of free scalars,
\begin{align}\label{limitcreat2N4}
{\cal O}(n)
&=
\int {d^3 \vec k \over 2 k^0 (2 \pi)^3} \sum_{\rm colors}
\left( a_1^{\dagger} (k) a_1 (k) - a_2^{\dagger} (k) a_2 (k) \right) \left({1 \over k^0} \delta^{(2)} (\Omega_{\vec n} - \Omega_{\vec k})  \right).
\end{align}
The only difference between this relation and \re{limitcreat2} is that Eq.\ \re{limitcreat2N4} takes into account the contribution of different scalar species
in $\mathcal{N}=4$ SYM. The type of scalars and their relative weights are controlled by the matrix $S$ introduced in \re{detectorsS}. In a similar
manner, the second flow operator $\mathcal{O}(n')$, corresponding to the matrix $S'$ in \re{detectorsS}, is given by \re{limitcreat2N4} with the indices
enumerating the scalars changed from 1 and 2 to 3 and 4, respectively.

In the Born approximation, the source \re{state} creates a pair of free scalars out of the vacuum. Their contribution to the weighted cross section is
\begin{align}
\label{ampweight0}
\vev{ {\cal O} (n) {\cal O} (n') }_{\bf 105}
=
\sigma_{\rm tot}^{-1}
\int  \text{dPS}_{2}   \big|\vev{s_1(k_1) s_3(k_2) |  O(0)|0} \big|^2
(k_1^0\, k_2^0)^{-1} \,  \delta^{(2)}(\Omega_{\vec k_1}-\Omega_{\vec n})\delta^{(2)}(\Omega_{\vec k_2}-\Omega_{\vec n'})
\, ,
\end{align}
where $\vev{s_1(k_1)s_3(k_2) |O(0)|0}$ is the transition amplitude into a two-particle final state $\ket{s_1, s_3}$. Here the subscripts indicate
the type of the scalars $s_I$ (with $I=1,\dots,6$). These are the only ones that satisfy the following two criteria: they are created by the operator \re{state}; they
are detected by one of the detectors \re{limitcreat2N4}. The square of the amplitude is a constant,
\begin{align}
\label{scattamplx0}
|\vev{s_1 (k_1) s_3 (k_2) |  O(0)|0}|^2 &=  2 \pi^4
\, ,
\end{align}
that depends on the normalization of the source operator, see Eq.\ \re{state}. Here and below, we use the conventional notation for the
Lorentz invariant phase space measure for $\ell$ massless particles with total momentum $q^\mu$,
\begin{align}
\int \text{dPS}_{\ell}  = \int  {\prod_{i=1}^\ell {d^4 k_i\over (2\pi)^4}\,2\pi\delta_+(k_i^2)} \, (2\pi)^4 \delta^{(4)}\big(q-\sum_{i=1}^\ell k_i\big)
\, ,
\end{align}
with $\delta_+(k^2)=\theta(k^0)\delta(k^2)$.
After an elementary calculation, Eq.\ \re{ampweight0} yields a result in complete agreement with  Eqs.\ \re{oneandtwo} and \re{BornF}.

As we come to the consideration of $O (a)$ effects, we shall focus on the region $0<z<1$ away from the kinematical boundaries. This leads to significant simplifications in the calculations. Namely, we can safely neglect the virtual corrections to the Born amplitude \re{scattamplx0} and take into account only the production of an extra gluon
in the two-scalar final state. The relevant scattering amplitude squared is
\begin{align}
\label{scattamplx1}
\big|\vev{s_1(k_1) s_3(k_2)g (k_3) |  O(0)|0} \big|^2 = 32 \pi^6 a \frac{s_{12}}{s_{13}s_{23}}
\, ,
\end{align}
where we used the fact that all particles are on-shell, $k_i^2 = 0$, and the Mandelstam invariants are denoted by $s_{i j} = (k_i + k_j)^2$.
Then, the order $O(a)$ correction to the double scalar flow correlation takes the form
\begin{align}
\label{ampweight}
\la {\cal O}(n) {\cal O}(n') \ra &=  \sigma_{\rm tot}^{-1}
\int   \text{dPS}_{3}   \big|\vev{s_1(k_1) s_3(k_2)g (k_3) |  O(0)|0} \big|^2
(k_1^0\, k_2^0)^{-1} \,
\delta^{(2)}(\Omega_{\vec k_1}-\Omega_{\vec n})
\delta^{(2)}(\Omega_{\vec k_2}-\Omega_{\vec n'})
\, .
\end{align}
In Appendix \ref{WeightedAmplitudes} we show that this expression reproduces the order $O(a)$ result \re{weakcouplCHF} of the correlation function computation 
for $0 < z < 1$. A similar calculation for the double energy flow was also performed in Ref.\ \cite{Engelund:2012re} and we find agreement with the functional 
$z$-dependence reported in that work.

The calculation of the  scalar flow correlation as a weighted cross section is rather simple at one loop, however it becomes very hard already 
at next-to-leading order. One needs to deal with
tree amplitudes with production of
four particles and one-loop amplitudes with three particles in the final state. While the corresponding transition amplitudes are available 
in the literature \cite{Brandhuber:2011tv,Bork:2011cj},
at this order the calculation requires an IR regulator, making it very involved. On the other hand, as we already mentioned before, the correlation function approach
bypasses these complications with ease. In addition, as we demonstrate in the next subsection, it allows us to compute the scalar flow correlations at strong coupling, a regime
unattainable in the amplitude approach.

\subsection{Double scalar flow correlation at strong coupling}

According to the  AdS/CFT correspondence \cite{Maldacena:1997re,Gubser:1998bc,Witten:1998qj}, the Euclidean correlation 
function of four half-BPS operators in planar $\mathcal N=4$
SYM at strong coupling is equal to the four-point scattering amplitude of massless scalars in the supergravity approximation to type 
IIB string theory on an AdS$_5 \times$S$^5$ background.
For the strong coupling counterpart of the correlation function \re{weakcoupl}, the result is \cite{Arutyunov:2000py,Arutyunov:2000ku,Eden:2000bk}
\begin{align}
\label{Strong4point}
&
\vev{O^{\dagger}(x_1) \widetilde O(x_2) \widetilde O (x_3) O(x_4)}_c
= \frac{1}{(2 \pi)^4} {1 \over x_{12}^2 x_{24}^2 x_{13}^2 x_{34}^2} \\
&\qquad\qquad\qquad
\times
\Bigg\{
{1\over 2}
+
\frac{1}{\pi^2} x_{12}^2 x_{23}^2 x_{34}^2 x_{14}^2
\left[
\frac{3}{2} D_{2222} + \left( \frac{v}{u} + \frac{1}{u} - 1 \right) x_{12}^2 D_{3322} - \frac{1}{2 x_{34}^2} D_{2211}
\right]
\Bigg\}
\, , \nonumber
\end{align}
where  the subscript $c$ indicates  that only the connected contribution is considered. The disconnected one, given by a product of two-point correlation functions, is protected from quantum corrections. The first term in the curly brackets formally coincides with the Born approximation (see Eq.\
\re{weakcoupl}) whereas the second term involves $D-$functions whose definition is recalled in Appendix \ref{Dfunctions}. The indices of the $D-$functions indicate the conformal weights
at the corresponding four points $x_1,\dots,x_4$. Though it is not explicit in the above representation, the second line in \re{Strong4point} is actually a function of the conformal cross-ratios
$u$ and $v$ only. This can be seen by converting it to the Mellin form. The main steps are reviewed in Appendix \ref{Dfunctions}. The result has the general form \re{weakcoupl} with the strong-coupling
Mellin amplitude given by the following remarkably simple expression
\begin{align}\label{strcoupla}
M(j_1,j_2) =
- \left[\Gamma(1-j_1) \Gamma(1- j_2) \Gamma(1+j_1 + j_2)\right]^2 \left(1 + j_1 + j_2 \over 2 j_1 j_2 \right)
\, .
\end{align}
The corresponding reduced Mellin amplitude \re{euclidmellinB1} takes the form
 \begin{align}\label{strcoupla1}
\widetilde M(j_1, j_2) = - {(1 + j_1 + j_2)(j_1 + j_2)^2 \over  2 j_1 j_2} \,.
\end{align}
We recall that relations \re{strcoupla} and \re{strcoupla1} are valid in planar $\mathcal N=4$ SYM at strong coupling.

Let us apply \re{weightedcross} and \re{K} to compute the {event shape function} resulting from Eq.\ \re{strcoupla}. As mentioned earlier, the first term in the curly brackets in \re{Strong4point} has 
already appeared at weak coupling and it produces $\mathcal{F}^{(0)} (z)$ from Eq.\ \re{BornF}. However, the  same contact term, but with an opposite sign, comes also from the second term in
\re{Strong4point}. As a result, ${\cal F}^{\rm (strong)}(z)$ is regular at $z=1$ and is given by the following Mellin integral
\begin{align}\label{strcouplaB} \nonumber
\mathcal{F}^{\rm (strong)}(z)
&=
- {\pi \over 4} \int \frac{dj_1 \, dj_2}{(2 \pi i)^2} {(j_1+j_2)^2(1+j_1 + j_2) \over j_1 j_2 \sin \pi (j_1 + j_2)} \left( {1-z \over z} \right)^{j_1 + j_2 -1}
\\
&\qquad\qquad\qquad\qquad\quad \  \
=
{1 \over 4} \sum_{k=1}^\infty k (1 + k)  \lr{z - 1\over z}^{k-1}
= {z^3 \over 2} \,,
\end{align}
where we first shifted one of the integration variables, e.g., $j_2 \to j_2 - j_1$, and then evaluated the resulting integrals by closing the contour and picking the
contributions from the residues. The same result \re{strcouplaB} can be obtained from the discontinuity representation \re{lastFourier} using the expression
for the function ${\cal G}(\gamma)$ at strong coupling
\begin{align}
{\cal G}^{(\rm strong)}(\gamma)=  {1 \over 32 \pi^3} {1 - \gamma^2 + 2 \gamma \ln \gamma \over \gamma (1 - \gamma)^3} \,,
\end{align}
which we find substituting \re{strcoupla} into \re{calG}. Thus, the double scalar flow correlation takes the form
\begin{align}\label{strcouplC}
\vev{ {\cal O} (n) {\cal O}(n') }_{\bf 105}^{\rm strong} = { 1 \over q^2 (nn')}  {z^3 \over 8 \pi^2}
\, .
\end{align}
Using the relation to the energy flow correlation \re{relation}, we get the following expression at strong coupling
\begin{align}\label{strcouplD}
\la  {\cal E} (n)  {\cal E} (n') \ra^{\rm strong} = { q^8 \over 16 \pi^2 (qn)^3 (qn')^3}
\, .
\end{align}
In the rest frame of the source $q^\mu = (q^0,0,0,0)$, it reproduces the result of Ref.\ \cite{Hofman:2008ar}, namely,  that  at strong coupling the energy is distributed homogeneously in the final 
state (see also \cite{Mul10}). This serves as a nontrivial check of the correlation function approach.

The following comments are in order. Notice that relation \re{relation} allows us to predict the leading correction to \re{strcouplC} at strong coupling
by making use of the corresponding computation for the energy flow \re{strcouplD} reported in \cite{Hofman:2008ar}
\begin{align}\label{strcouplBE}
\vev{ {\cal O} (n) {\cal O} (n') }_{\bf 105}^{\rm strong} = { 1 \over q^2 (nn')}  {z^3 \over 8 \pi^2} \left[1 + {1 \over a} (6 z^2 - 6 z + 1) \right]
\, .
\end{align}
It would be natural to expect to find an oder $O(1/\sqrt{a})$ correction to the supergravity approximation, associated with the first term in  the $\alpha^\prime$-expansion, but
this actually vanishes due to a subtle interplay in the string amplitude (see Eq.\ (4.5) in Ref. \cite{Hofman:2008ar}). Thus, the leading $\alpha^\prime$-effect intervenes only at the next order, as shown in Eq.\ \re{strcouplBE}. Matching this relation with the Mellin representation \re{weightedcrossB}, we obtain a prediction for the first string correction to the integrated
reduced Mellin amplitude,
\begin{align}\label{strcouplBE2}
\int_{- \delta - i \infty}^{- \delta + i \infty} {d j_1 \over 2 \pi i} \widetilde M(j_1 , j_2 - j_1 )
=
{1 \over 2} j_2 (1 + j_2) + {1 \over 8 a}  (j_2 (1 + j_2))^2 + O (a^{-3/2}) \, .
\end{align}

In contrast with the correlation function approach, the calculation of \re{strcouplC} and \re{strcouplD} at strong coupling using amplitudes is highly problematic. 
The weighted cross section
\re{event-shapes} is given by an infinite sum of transition amplitudes $\bra{X} O \ket{0}$,  to be evaluated at strong coupling. According to the AdS/CFT correspondence, these transition
amplitudes are determined at strong coupling by the area of minimal surfaces in AdS space. The latter can be found using integrability techniques, the so-called $Y-$system
\cite{Alday:2007he, Alday:2010vh, Maldacena:2010kp, Gao:2013dza}. This leads to the exponential fall-off $\bra{X} O \ket{0} \sim \e^{- \sqrt{a} \,{\rm Area}(X)}$ for
any number of particles in the final state $\ket{X}$. Thus, in order to calculate the event shapes \re{event-shapes}  at leading order at strong coupling, we need to know the
explicit solution of the $Y-$system for any final state $\ket{X}$ and then to find a way to  resum an infinite number of contributions on the right-hand side of \re{event-shapes}.  These
complications make the use of scattering amplitudes for computing the charge flow correlations at strong coupling extremely hard.

Next, let us comment on the importance of stringy effects for the charge flow computations.
These can have profound implications for the event shapes, arising from their sensitivity to the Regge asymptotics of the string scattering amplitudes,
as was emphasized\footnote{See the discussion around Eqs.\ (4.20) and (4.22) in Ref.\ \cite{Hofman:2008ar}.} in Ref.\  \cite{Hofman:2008ar} and as we recall below.
Via the AdS/CFT correspondence, the double charge flow correlation is related to a four-point string scattering amplitude.
 As explained in \cite{Hofman:2008ar}, the relevant AdS computation can be approximated by the flat space one. This simplifies the task of
computing string corrections to the double charge flow  since one can use the flat-space Shapiro-Virasoro amplitude. It is then clear that the double flow
correlation can be sensitive to the high-energy behavior of the latter. Namely, according to \cite{Hofman:2008ar} one of the integrals over the detector working
time is translated into an integral over the Mandelstam variable $s$, while the second one is removed by the energy-momentum conserving delta function. The upper limit of $s$
extends all the way to infinity and, thus, the result potentially depends on the large-$s$ asymptotics of the integrand that involves the scattering amplitude as a multiplicative factor.
It is well known that string amplitudes display the Regge behavior $\exp(\alpha' t \ln s)$ versus the power law fall-off of gravity amplitudes as $s \to \infty$.
This may lead to an order-of-limits non-commutativity for the operations of time integration and taking the string tension to infinity,
$1/\alpha' \to \infty$. The supergravity approximation used in our analysis corresponds to performing the latter step first. It leads to the decoupling of all heavy string modes before
the time integral is performed. Indeed, it was noticed in Ref.\ \cite{Hofman:2008ar} that in certain circumstances, the result for some charge flow correlations depends on the
order of limits. The details of this phenomenon depend on the detectors at hand. For example, for the energy flow correlations it was noticed that the problem does not emerge
while for the R-charge flow correlations the operations cease to commute. In this case the premature decoupling of stringy modes yields divergences. The lesson that one
learns is that the computation in the supergravity approximation may lead to unphysical divergences which do not appear if one first performs the computation in the
full string theory and then sends $\alpha' \to 0$ only at the very end. By considering different detectors and states and repeating the steps of \cite{Hofman:2008ar}, it becomes
clear that the particular scenario of inducing divergences does not seem to take place in general. Thus, one can expect that for certain configurations of the R-charge detectors and
 sources the divergence will not appear. This is exactly the case that we discussed in this section. However, as we  demonstrate in \cite{paper2}, other channels will be plagued by
 these problems and thus their elimination will require a stringy calculation.

Notice that one should distinguish between different sources of potential divergences that could contaminate the charge flow correlations. Some of them are the familiar IR
divergences, while others are spurious and appear due to the effect described above. The detailed consideration of this phenomenon 
in the context of ${\cal N}=4$ SYM requires further
analysis and will be presented elsewhere.

\subsection{Constraints at finite coupling}\label{sect-con}

As announced in Eq.\ \re{relation}, the double scalar flow correlation is related to the energy flow 
correlation in $\mathcal N=4$ SYM. Based on the physical interpretation of the latter, we
expect that, for any value of the coupling constant $a$ and for finite $N_c$,\footnote{For recent progress on ${\cal N}=4$ SYM correlation functions at finite coupling,
see \cite{Honda:2013nfa}.}  they have to obey the conditions of {\it IR safety} (finiteness), {\it positivity} and {\it regularity}. Exploiting
relation \re{relation}, these conditions can be cast as certain predictions for the four-point correlation function of half-BPS operators 
\re{weakcoupl}, or equivalently for the corresponding Mellin amplitude $M(j_1,j_2)$.

As discussed in Section \ref{LimitTime}, for the scalar flow correlation  \re{weightedcrossB} to be IR finite, the reduced Mellin amplitude $\widetilde M(j_1 , j_2 - j_1)$ should decay at
large $j_1$ as $1/j_1$ or faster, see Eq.~\re{irsafety}. Notice that there are  {\it a priori} no reasons to expect that this correlation should be finite in a generic CFT.
In the case of ${\cal N}=4$ SYM, this condition follows from the relation \re{relation}, since the energy flow correlations are known to be IR safe.  Indeed, using the 
obtained expressions at weak and strong coupling, Eqs.~\re{OneLoopMellinAmp} and \re{strcoupla1}, respectively, we verify that the Mellin amplitude scales at 
large $j_1$ as $\widetilde M(j_1 , j_2 - j_1 ) \sim 1/ j_1^2$, in agreement with \re{irsafety}. It would be very interesting to understand the physics of the limit \re{irsafety} 
and what properties of the correlation functions and/or of the CFT govern the large $j_1$ behavior.

Turning to the second condition, we recall that the positivity of the energy flow correlations leads, through \re{relation}, to an analogous property of the
{event shape function} \re{oneandtwo}, ${\cal F}(z) \geq 0$ for $0\le z\le 1$. Making use of relation \re{weightedcrossB} and
introducing the variable $\eta = z/(1-z)$, we can invert the Mellin transformation in \re{weightedcrossB} to get
\begin{align}\label{inversemellin}
\int_0^{\infty} d \eta \, \eta^{j_2} {\cal F} \big(\eta/(1+\eta)\big)
=
\frac{\pi}{2 \sin(\pi j_2)} \int_{- \delta - i \infty}^{- \delta + i \infty}
\frac{d j_1}{2 \pi i} \widetilde{M} (j_1, j_2 - j_1)
\, .
\end{align}
Choosing $j_2$ in \re{inversemellin} to be real and such that the $\eta$-integral on the left-hand side converges, we immediately conclude that the expression 
in the right-hand side should be non-negative,
\begin{align}\label{positivity}
\frac{1}{ \sin(\pi j_2)}\int_{- \delta - i \infty}^{- \delta + i \infty} {d j_1 \over 2 \pi i} \widetilde M(j_1, j_2 - j_1) \ge 0
\, ,
\end{align}
with $j_2$ constrained as above. One can easily check that both the weak and strong coupling results, Eqs.~\re{OneLoopMellinAmp} and \re{strcoupla1}, 
respectively, obey the constraint \re{positivity}. It would be interesting to understand whether the positivity property \re{positivity} leads to new constraints 
on the three-point functions and scaling dimensions of conformal operators in ${\cal N}=4$ SYM at finite coupling or whether it is automatically satisfied once 
the known unitarity bounds are fulfilled.\footnote{In the particular case of extremal ${a / c}$, the positivity of the multiple-detector  energy correlations can 
be used to fix all of them \cite{Zhiboedov:2013opa}.}

Last but not least, another property of the energy flow correlations that is believed to hold at finite coupling is their {\it regularity}. By this we mean the absence
of terms like $\delta(1-z)$ which appear at any finite order in the weak coupling expansion of $\mathcal F(z)$. Indeed, as we observed above, 
the energy flow correlations are completely regular at strong coupling, see Eq.\ \re{strcouplD}. Making use of Eq.\ \re{lastFourier}, we see that the term 
$\delta(1-z)$ in $\mathcal F(z)$ is related to the $1/\gamma-$pole in the function $\mathcal G(\gamma)$. Thus, the contribution from 
\re{doubledisc} proportional to $\delta(1-z)$ takes the form
\begin{align}\label{regularity}
 \frac{1}{2 \pi^2} \delta(1-z) \int_{- \infty}^{0} d w  \int_{1}^{\infty} d \bar w   {w - \bar w \over w \bar w (1-w)(1 - \bar w)}
{\rm Disc}_{w} {\rm Disc}_{\bar w}  \Phi(w, \bar w; a)
\, .
\end{align}
Naively, it seems that the regularity of the energy flow correlations and of the matching scalar flow observables (upon using Eq.\ \re{relation})
implies that the integral in \re{regularity} should vanish. However, at strong coupling we observed that the assumption
about the Mellin transform of the connected correlator being a meromorphic function breaks down. The reason for this is the 
presence of the connected Born-like contribution (the first term inside the
curly brackets in \re{Strong4point}). The latter
is clearly not captured by Eq.\ \re{regularity} which assumes the existence of a meromorphic Mellin amplitude, Eq.~\re{disc}. Thus,   relation \re{regularity} has to be modified by taking
into account  the additional contribution $\frac14 \delta(1-z)$ (see Eq.~\re{BornF}).
In this way, we arrive at the following relation  
\begin{align}\label{regularityB}
&\frac14+{1\over 2\pi^2}\int_{- \infty}^{0} d w  \int_{1}^{\infty} d \bar w   {w - \bar w \over w \bar w (1-w)(1 - \bar w)}   {\rm Disc}_{w} {\rm Disc}_{\bar w}  \Phi(w, \bar w; a)=0\,.
\end{align}
Together with \re{M-imp}, this relation fixes the leading asymptotic behaviour of the Mellin amplitude $M(j_1, j_2)$ for small $j_1$ and $j_2$
\begin{align}
 M(j_1, j_2) \sim  - {1 \over 2 j_1 j_2} \,,\qquad \text{for $j_1,j_2\to 0$}\,.
\end{align}
We would like to emphasize that Eq.\ \re{regularityB} holds under the assumptions that for arbitrary coupling the connected part of the correlation
function can be separated into a sum of a Born-like term and a Mellin integral, as indicated in the second line of \re{weakcoupl}. In addition, the 
double discontinuity ${\rm Disc}_{w} {\rm Disc}_{\bar w}  \Phi(w, \bar w; a)$ should vanish at the end points of the cuts, at $w = 0$ and $\bar{w} = 1$,
so that the integral in \re{regularityB} converges around these points.

This implies in particular that the relation \re{regularityB} does not hold at any finite order in the weak coupling expansion. Indeed, it is easy
to see from \re{Phi1} that the double discontinuity does not vanish at the end points of the cuts already at one loop. This leads to
a double logarithmic divergence of the integral in \re{regularityB}. We anticipate that, upon resummation to all orders in the coupling,
such divergences exponentiate and, as a consequence, produce a vanishing contribution to \re{regularityB}.

It is straightforward to verify that both assumptions formulated above hold at strong coupling. In this case, the double discontinuity
${\rm Disc}_{w} {\rm Disc}_{\bar w}  \Phi(w, \bar w; a)$ can be found with the help of \re{disc} and \re{strcoupla}.
Due to the simple relation between the Mellin amplitudes at weak and strong coupling, Eqs.~\re{OneLoopMellin} and \re{strcoupla}, respectively,
the result can be expressed in terms of the double discontinuity at one loop, Eq.~\re{Phi1},
\begin{align}\notag \label{strongregul}
 {\rm Disc}_{w} {\rm Disc}_{\bar w}  \Phi^{\rm(strong)}(w, \bar w)
 &=   2(u\pa_u) (v \pa_v)  \left( u \pa_u + v \pa_v + 1 \right) {\rm Disc}_{w} {\rm Disc}_{\bar w}  \Phi^{(1)}(w, \bar w)
 \\
& =  {\pi^2 \over 2} (u\pa_u) (v \pa_v)\left( u \pa_u + v \pa_v + 1 \right)  {1 \over \sqrt{(1+u+v)^2 - 4 u v}}
\, ,
\end{align}
with $u = - w \bar w$ and $v = - (1-w)(1 - \bar w)$. Substituting this expression into the left-hand side of \re{regularityB}, we find after some algebra that the 
relation  \re{regularityB} is indeed satisfied.  If the above assumptions continue to hold for the subleading (stringy) corrections to the function 
$\Phi(w, \bar w, a) = \sum_{k=0}^{\infty}  {a^{-k/2} } \Phi^{\rm (strong)}_{k}(w,\bar w)$, each term in the strong coupling expansion 
$\Phi^{\rm (strong)}_{k}(w,\bar w)$ (with $k>0$)  should integrate to zero upon its substitution into \re{regularityB}.

\section{Summary and conclusions}

In this paper, we presented a new approach to computing event shape distributions or, more precisely, charge flow correlations in 
a generic conformal field theory. These observables are
familiar from collider physics studies and they describe the angular distribution of global charges in outgoing radiation created from 
the vacuum by some source. To simplify our discussion, we focused on
the case where both the source and the charge flow operators are expressed in terms of scalar operators of the same scaling dimension. 
In this case, the charge flow correlations are given
by the Wightman correlation function $G_W$ of scalar operators in a certain (detector) limit, which includes sending some of the 
operators to future null infinity and integrating over their
position on the light front \re{detector11}, \re{chargeflowcorr}. We then explained how to compute the same quantity using the 
Euclidean counterpart $G_E$ of the Wightman correlation function
$G_{W}$. The two functions, $G_W$ and $G_E$, are related to each other through a nontrivial analytic continuation which, in the 
framework of CFT, is especially easy to perform using the
Mellin representation of the correlation function.

We demonstrated that in the particular case of the double flow correlations, the symmetries of the problem fix  the result up to an 
arbitrary  single-variable event shape function \re{oneandtwo}. We
explained that the event shape function is unambiguously fixed by the Euclidean four-point correlator of the scalar operators or, 
more precisely, by its Mellin amplitude. Namely, the two-point correlations
are given by a convolution of the Mellin amplitude with the so-called detector kernel \re{K}, which is a universal meromorphic 
function independent of the CFT. The same result can be reformulated
directly in configuration space, without going to the Mellin representation, as a certain Lorentzian double discontinuity of 
the Euclidean correlation function integrated along the cuts,
Eqs.~\re{lastFourier} and \re{doubledisc}. Our analysis can be extended to correlations involving conserved currents and/or 
stress tensor. It goes through exactly the same steps as before, even
though it is technically more involved due to the complicated tensor structure of the correlation functions.

To illustrate the general formalism, we computed the double flow correlations in ${\cal N} = 4$ SYM from the four-point 
correlation function of half-BPS scalar operators,
which have been studied thoroughly both at weak and strong coupling. One of the remarkable features of this theory is 
that the scalar and energy flow correlations are related to each other
as in Eq.\ \re{relation} (see \cite{paper2}).\footnote{This relation is based  solely on $\cN=4$ superconformal symmetry, it 
does not rely on the dynamics of the theory.} At weak coupling, we verified that
our approach leads to expressions for the even shape function that agree with the result obtained using the conventional 
amplitude approach. At strong coupling, we reproduced the
finding of Ref.~\cite{Hofman:2008ar} starting from the four-point correlation function obtained via AdS/CFT in the supergravity 
approximation \cite{Arutyunov:2000py}. These two tests serve as an excellent check of the formalism.

Using the aforementioned relation between the scalar and energy flow correlations in ${\cal N} = 4$ SYM, we translated the first string correction to the energy
flow correlations, computed in \cite{Hofman:2008ar}, into a prediction,  Eq.\ \re{strcouplBE}, for the $1/a-$correction to the Mellin amplitude 
\re{strcouplBE2} corresponding to the four-point correlation function of half-BPS operators. In addition, we formulated conditions on the Mellin amplitude 
that follow from the infrared finiteness, positivity and regularity of the energy flow correlations.  Most 
importantly, they should be valid in ${\cal N} = 4$ SYM for arbitrary coupling and away from the planar limit.

One of our motivations has to do with developments in QCD. The existing methods for computing energy flow correlations rely on amplitudes and are not well tailored to obtain
analytic predictions beyond  leading order of perturbation theory. The current approach offers a possibility to avoid the inefficiency of this 
standard method by exploiting the results for the
correlation functions available in the literature. As we  show in \cite{paper3}, in the special case of $\mathcal N=4$ SYM the energy flow correlations can be computed
analytically at weak coupling to two loops, at the very least. Based on various examples of observables previously studied in this theory, we may 
speculate that the ${\cal N} = 4$ SYM result  describes the
`most complicated' part of the QCD expressions.

Let us point out that the underlying concepts of the current approach go far beyond their phenomenological QCD applications. 
The study of energy flow correlations in CFTs is of importance on its own since
they describe second-order phase transitions, end points of renormalization group flow and quantum gravities in AdS.  Imposing 
natural physical conditions on the energy flow correlations (finiteness,
positivity and regularity) should lead to additional constraints on the correlation functions in unitary CFTs. For example, in the 
case of one-point correlation, the positivity condition $\vev{\mathcal E(n)}
\geq 0$ is known to be related to the violation of causality for perturbations propagating on top of nontrivial gravitational 
backgrounds \cite{Brigante:2008gz, Hofman:2009ug, Buchel:2009sk, Camanho:2009vw}.
Studying these conditions is especially simple in CFTs possessing additional symmetries such as $\mathcal N=4$ SYM, 
since the correlation functions in such theories are highly constrained. In particular,
it would be interesting to understand if we can use them as  additional input in  recent studies of ${\cal N}=4$ SCFT 
bootstrap \cite{Rattazzi:2008pe,Beem:2013qxa}. Next, the charge flow correlations
possess another attractive feature  that can be seen through the AdS/CFT correspondence. As was shown in \cite{Hofman:2008ar}, 
they are sensitive to the Regge limit of scattering amplitudes in the dual theory.
Due to this fact, for a general choice of the detectors, their computation in the supergravity approximation yields a divergent 
result \cite{paper2}. The divergences are unphysical and can be subsequently
cured by going beyond the supergravity limit and by taking into account  contributions of heavy string modes. It 
would be interesting to better understand the underlying mechanism since it is
related to the emergence of locality in AdS.

\section*{Acknowledgments}

We would like to thank Vladimir Braun, Simon Caron-Huot, Juan Maldacena and Raymond Stora for interesting discussions. We are indebted to Juan Maldacena for careful reading 
of the manuscript and useful comments. The work of A.B.\ was supported by the U.S.\ National Science Foundation under the grant No.\ PHY-1068286 and CNRS. G.K.\ and E.S.\ acknowledge 
partial support by the French National Agency for Research (ANR) under contract StrongInt (BLANC-SIMI-4-2011). A.Z.\ was supported in part by the U.S.\ National Science
Foundation under Grant No.\ PHY-0756966. A.B., S.H. and A.Z.\ are grateful to the Institut de Physique Th\'eorique (Saclay) for the hospitality extended to them at various stages of
the project.

\appendix

\section{Scalar detectors in CFT}
\label{DdimCFT}

In this appendix, we present the calculation of the double-detector correlation $\la {\cal O}({n})   {\cal O}({n}') \ra_{q}$ in the general case of a $D-$dimensional
CFT
\begin{align}\label{app1}
\la {\cal O}({n})   {\cal O}({n}') \ra_{q} = \lr{\sigma_{\Delta}(q)}^{-1} \int d^D x_1 \e^{iqx_1} \vev{0|O^\dagger (x_1){\cal O}(n){\cal O}(n')O(0) |0}\,.
\end{align}
We assume for simplicity that the source is a conformal primary scalar operator $O(x)$ with  arbitrary conformal weight $\Delta$, normalized as
\begin{align}\label{app2}
\vev{0|O^\dagger(x) O(0)|0} = {1\over (-x^2+i\epsilon x^0)^\Delta}\,,
\end{align}
and define the scalar flow operator as
\begin{align}\label{app3}
{\cal O}(n) &= (n \bar n)  \int_{- \infty}^{\infty} d x_{-} \lim_{x_{+} \to \infty} x_+^\Delta\, \widetilde{O} \left( x_+ n + x_- \bar n \right)\,.
\end{align}
As before, the overall normalization factor is given by the Fourier transform of the two-point Wightman correlation function
\begin{align}\label{app4}
\sigma_{\Delta}(q) = \int { d^D x \e^{iqx} \over (-x^2+i\epsilon x^0)^{\Delta}}  =
 \theta(q^0)\theta(q^2) {2\pi^{D/2+1}(q^2/4)^{\Delta-D/2}  \over \Gamma(\Delta)\Gamma(\Delta+1-D/2)}\,.
\end{align}
In what follows we do not display the theta-function and assume that the conditions $q^0>0$ and $q^2>0$ are satisfied.

Substituting \re{app3} into \re{app1}, we can express $\la {\cal O}({n})   {\cal O}({n}') \ra_{q}$ in terms of  the four-point Wightman correlation
function of the scalar operators. In a close analogy with \re{G4-W}, we use the Mellin representation for this function
\begin{align}\notag
\vev{0|O^\dagger(x_1) \widetilde{O}(x_2) \widetilde{O}(x_3) O(x_4)|0}  = \int_{- \delta - i \infty}^{- \delta + i \infty} {d j_1 d j_2 \over (2 \pi i)^2}  M(j_1, j_2) (\hat x_{23}^2)^{-j_1-j_2} (\hat x_{14}^2)^{-j_1-j_2}
\\[2mm] \times
(\hat x_{12}^2)^{j_1-\frac{\Delta}2}( \hat x_{34}^2)^{j_1-\frac{\Delta}2}   (\hat x_{13}^2)^{j_2-\frac{\Delta}2} (\hat x_{24}^2)^{j_2-\frac{\Delta}2} \,,
\end{align}
where $\hat x_{ij}^2 = -x_{ij}^2 + i \epsilon x_{ij}^0$ and $x_{ij}\equiv x_i-x_j$.
Going along the same lines as in Sect.~2.3, we get
\begin{align}\notag
\la {\cal O}({n})   {\cal O}({n}') \ra_{q} & = \ {(2i)^{-2\Delta}\over \sigma_{\Delta}(q)}  \int_{- \delta - i \infty}^{- \delta + i \infty} {d j_1 d j_2 \over (2 \pi i)^2}  M(j_1, j_2)
\\ \notag
& \times \int d^D x_1\e^{iqx_1} (x_1^2-i\epsilon x_1^0)^{-j_1-j_2}   ((nn')/2)^{-j_1-j_2}
\\\notag
& \times\int dx_{2-} ((nx_1)-x_{2-}-i\epsilon)^{j_1-\Delta/2}(x_{2-}-i\epsilon)^{j_2-\Delta/2}
\\
& \times\int dx_{3-} ((n'x_1)-x_{2-}-i\epsilon)^{j_2-\Delta/2}(x_{3-}-i\epsilon)^{j_1-\Delta/2}\,.
\end{align}
The integration over the detectors time can be easily	performed with the help of the identity
\begin{align}
\int_{-\infty}^\infty dx_{2-} ((nx_1)-x_{2-}-i\epsilon)^{-a}(x_{2-}-i\epsilon)^{-b}
 = 2\pi i ((nx_1)-i\epsilon)^{1-a-b} {\Gamma(a+b-1)\over \Gamma(a)\Gamma(b)}\,.
\end{align}
Notice that the condition for the integrals over $x_{2-}$ and $x_{3-}$ to be convergent leads to a restriction on the
choice of the integration contour in the Mellin integral,  $\Re(j_1+j_2) < \Delta-1 $. In this way, we arrive at
\begin{align}\notag\label{appF}
\la {\cal O}({n})   {\cal O}({n}') \ra_{q} & = {(2i)^{2-2\Delta}\pi^2\over \sigma_{\Delta}(q) }  \int_{- \delta - i \infty}^{- \delta + i \infty} {d j_1 d j_2 \over (2 \pi i)^2}  M(j_1, j_2)
\lr{(nn')\over 2}^{-j_1-j_2}\left[\Gamma(\Delta-1-j_1-j_2) \over \Gamma(\Delta/2-j_1)\Gamma(\Delta/2-j_2)\right]^{2}
\\[2mm]
& \times   \int d^D x_1\e^{iqx_1} (x_1^2-i\epsilon x_1^0)^{-j_1-j_2}[((x_1n)-i\epsilon)((x_1n')-i\epsilon)]^{j_1+j_2+1-\Delta}\,.
\end{align}
Using the Schwinger parametrization
\begin{align}
[((x_1n)-i\epsilon)((x_1n')-i\epsilon)]^{-a}= {i^{2a}\over \Gamma^2(a)} \int_0^\infty d\omega d\omega' (\omega \omega')^{a-1} \e^{-i\omega(x_1n)-i\omega' (x_1n')}\,,
\end{align}
we can perform the Fourier integration in \re{appF} with the help of identity \re{app4}. Namely, the Fourier integral in the second line of \re{appF} can be rewritten  as
\begin{align}
{i^{2(\Delta-1)}\over \Gamma^2(\Delta-1-j_1-j_2)}\int_0^\infty d\omega d\omega' (\omega \omega')^{\Delta-2-j_1-j_2} \sigma_{j_1+j_2}(q-n \omega-n'\omega')\,,
\end{align}
where $\sigma_{j_1+j_2}(q)$ is given by \re{app4} with $\Delta$ replaced with $j_1+j_2$. Its evaluation is straightforward and leads to a ${}_2 F_1-$hypergeometric function. Finally, we combine together various factors in \re{appF} and obtain
\begin{align}\notag\label{app-fin1}
\la {\cal O}({n})   {\cal O}({n}') \ra_{q}  =   \pi^2  & {(q^2/4)^{\Delta-2}\over ((qn)(qn'))^{\Delta-1}}    \int_{- \delta - i \infty}^{- \delta + i \infty} {d j_1 d j_2 \over (2 \pi i)^2}  M(j_1, j_2)
\\[2mm] & \times \notag
z^{-j_1-j_2}{}_2 F_1 \bigg({\Delta-1-j_1-j_2,\Delta-1-j_1-j_2 \atop 2\Delta-1-j_1-j_2-{D\over 2}}  \bigg| z \bigg)
\\ & \times
 {\Gamma(\Delta)\Gamma(1+\Delta-{D\over 2}) \Gamma^2(\Delta-1-j_1-j_2)\over \Gamma(j_1+j_2)\Gamma^2({\Delta\over 2}-j_1)\Gamma^2({\Delta\over 2}-j_2)\Gamma(2\Delta-1-j_1-j_2-{D\over 2})}\,,
\end{align}
where $z=q^2 (nn')/(2 (qn)(qn'))$. Using the Barnes formula for the hypergeometric function, this can also be rewritten as
\begin{align}\notag\label{app-fin12}
\la {\cal O}({n})   {\cal O}({n}') \ra_{q}
=
\pi^2 &  {(q^2/4)^{\Delta-2}\over ((qn)(qn'))^{\Delta-1}} \Gamma(\Delta) \Gamma(1+\Delta- \ft{D}{2})
\int_{- \delta - i \infty}^{- \delta + i \infty} {d j_1 d j_2 d s \over (2 \pi i)^3} \widetilde M(j_1, j_2) \Gamma(j_1+ j_2)
\\[2mm]
& \times
(-1)^s z^{-j_1-j_2 + s} {\Gamma(-s) \Gamma(\Delta-1-j_1-j_2 + s)^2 \over \Gamma(2\Delta-1-j_1-j_2-{D\over 2}+s)}\,,
\end{align}
where we substituted the generalization of Eq.\ \re{euclidmellinB1} to arbitrary $\Delta$ and $D$,
\begin{align}
M(j_1, j_2)=\left[\Gamma(\ft{\Delta}{2} -j_1) \Gamma(\ft{\Delta}{2} - j_2) \Gamma(j_1 + j_2)\right]^2  \widetilde M(j_1, j_2)\, .
\end{align}

In the special case $D=4$ and $\Delta=2$, which is relevant for the discussion in Section~\ref{LimitTime}, relation \re{app-fin1}  simplifies significantly,
\begin{align}\notag\label{app-fin2}
\la {\cal O}({n})   {\cal O}({n}') \ra_{q}  =    & {2 \pi^2 \over q^2(nn')}    \int_{- \delta - i \infty}^{- \delta + i \infty} {d j_1 d j_2 \over (2 \pi i)^2}  M(j_1, j_2)
z^{1-j_1-j_2} (1-z)^{j_1+j_2-1}
\\ & \times {  \Gamma(1-j_1-j_2)\over \Gamma(j_1+j_2)\Gamma^2(1-j_1)\Gamma^2(1-j_2)}
\, .
\end{align}
We would like to emphasize that relations \re{app-fin1} and \re{app-fin2} were obtained for scalar operators normalized according to \re{app2}.
If we change the normalization as $O(x) \to \rho\, O(x)$, the expressions on the right-hand side of \re{app-fin1} and \re{app-fin2} will acquire an additional factor of $\rho^2$.

\section{Event shape function at one loop}
\label{SingularTermsAppendix}

In this appendix we show how to obtain the one-loop event shape function \re{Z1Subs} starting from the correlation function \re{weakcoupl}. Compared to the main text, we will switch the
order of operations and first perform the Mellin transform \re{calG} followed by the Fourier integral \re{fourierrep}. To this end, we substitute the one-loop Mellin kernel \re{OneLoopMellin}
into \re{calG} and compute the resulting Mellin transform by shifting the integration variable $j_2 \to j_2 - j_1$ and then evaluating the factorized product of Mellin integrals using
Cauchy's theorem. This yields the expression for $\mathcal{G} (\gamma)$ quoted in Eq.\ \re{calGoneloop}. Then the one-loop {event shape function} is given by the Fourier transform
\begin{align}\label{singular}
\mathcal{F}^{(1)} (z)
=
- {q^2 \over 4 (2 \pi)^3}  \int d^4 x  {{\rm e}^{i q x} \over x^2- i \eps x^0}  {{\rm Li}_{2}({\gamma -1 \over \gamma}) - \zeta_2 \over \gamma}
\, ,
\end{align}
where $\gamma$ is defined in Eq.\ \re{gamma}. This is the starting point of the analysis which follows.

The evaluation of the term in Eq.\ \re{singular} proportional to $\zeta_2$ is analogous to that of the Born contribution
\re{BornF}. It produces
\begin{align}
\label{Zeta2inF1}
{\cal F}^{(1)}(z) =  - {\zeta_2 \over 4 } \delta(1-z)+\dots
\, ,
\end{align}
and the ellipsis stands for the contribution of the dilogarithm term in \re{singular}. An efficient way to compute the latter is by using the well-known integral representation
\begin{align}
\label{GenFuncPoly}
{\rm Li}_2(y)=\int_0^\infty \frac{dt\,t}{\e^t/y-1} \, ,
\end{align}
which is valid for all $y$ except for $y\ge 1$ on the real axis. This  is suitable for our purposes since
\begin{align}
y \equiv {\gamma - 1 \over \gamma} = {\vec{x}_{\perp}^{2} (nn') \over 2 ((xn) - i \eps) ((xn') - i \eps)}
\end{align}
is complex, as follows from the $i \eps$ prescriptions stemming from the analytic continuation.
Here we made use of the Sudakov decomposition of the four-dimensional vector $x^\mu$
\begin{align}
x = \alpha n + \beta n' + x_\perp
\, , \qquad
\a= \frac{(xn)}{(nn')}\, ,
\quad
\b= \frac{(xn')}{(nn')}
\, ,
\end{align}
where $x_\perp$ is an Euclidean two-dimensional vector orthogonal to the null-vectors $n$ and $n'$,   $(x_\perp n)=(x_\perp n')=0$. Further,
  $x^2=2 (nn') \alpha\beta-\vec x_\perp^2$ and the integration measure reads $ d^4 x = (nn') d\alpha d\beta d^2 \vec x_\perp$.
In this way the integral \re{singular} takes the form
\begin{align}\label{singularB}
 \int d^4 x  {{\rm e}^{i q x} \over x^2- i \eps x^0}  {1 \over \gamma} {\rm Li}_{2}(1 -1/\gamma)
 &
 =
{1 \over 2 (nn')^2} \int_0^\infty dt \, t \, e^{- t}   \int  d^2 \vec{x}_\perp \vec{x}_{\perp}^2 {\rm e}^{- i (\vec{q}_\perp \vec{x}_\perp)}
\\
&
\times
\int_{-\infty}^{\infty} d \beta {{\rm e}^{i (qn') \beta} \over (\beta - i \eps)^2}
\int_{-\infty}^{\infty} d \alpha {{\rm e}^{i (qn) \alpha} \over  (\alpha - i \eps) \left( \alpha - {\vec{x}_{\perp}^2 {\rm e}^{-t} \over 2  (nn') (\beta - i \eps)}- i \eps\right)}
\, , \nonumber
\end{align}
where $q_\perp$ is the two-dimensional transverse vector in the Sudakov decomposition of $q^\mu$,
\begin{align}
\label{Sudak}
q = {(qn) \over (nn')} n' + {(qn') \over (nn')} n + q_{\perp}
\, .
\end{align}

The integral over $\alpha$ in \re{singularB} can be evaluated by residues. Since both poles are located in the upper half-plane, the integral vanishes unless $(qn)>0$. So,  \re{singularB} gives
\begin{align}\label{singularC}
- 2 \pi i  {\theta(qn) \over (nn')} \int_0^\infty dt\,t  \int  d^2 \vec{x}_\perp \, {\rm e}^{- i (\vec{q}_\perp \vec{x}_\perp)}
\int_{-\infty}^{\infty} d \beta {{\rm e}^{i (qn') \beta} \over (\beta - i \eps)}
\left[ 1 - \exp \left( {i {(qn) \vec{x}_{\perp}^2 {\rm e}^{-t} \over 2  (nn') (\beta - i \eps)} } \right) \right]
\, .
\end{align}
The subsequent calculation involves first taking the integral over $x_\perp$ and then over $\beta$.
This produces  delta functions, so that the remaining $t-$integration becomes trivial,
\begin{align}\label{singularD}
(2 \pi)^3 {\theta(q^0) \theta(q^2) \over  (qn) (qn')} \int_0^\infty dt\,t \left[ \delta(1-z) - \delta(1 - z - e^{-t}) \right]
=
2 (2 \pi)^3 {\theta(q^0) \theta(q^2) \over q^2 (nn')} z \left[ {\log (1- z) \over 1-z}  \right]_+,
\end{align}
where we used the conventional definition for the plus-prescription
\begin{align}
\label{PlusPrescrip}
[ f(z) ]_{+} = f(z) - \delta(z-1) \int_0^1 d y f(y)
\, ,
\end{align}
for any test function $f(z)$. In the first term in the square brackets on the left-hand side of \re{singularD} we used the identity between tempered distributions,
\begin{align}
\delta^{(2)} (\vec{q}_\perp) =
(2\pi)^{-1} \delta (|\vec{q}_\perp|)/|\vec{q}_\perp|=
{1\over \pi} \delta ({\vec{q}_\perp}^{\,2}) = {1\over \pi q^2} \delta(1-z)\,,
\end{align}
valid for Schwartz test functions independent of the angular variable in the two-dimensional plane, and made use of the identity $(1-z)/z=q_\perp^2/q^2$,
see Eqs.\ \re{z} and \re{Sudak}.

Combining Eqs.\ \re{Zeta2inF1} and \re{singularD}, we find the result \re{Z1Subs} quoted in the body of the paper. Relation  \re{Z1Subs}  differs from \re{weakcouplCHF} by terms
proportional to $\delta(1-z)$. Such terms are needed
to restore the finiteness of the moments of the event shape function,
\begin{align}\label{mom}\notag
\int_0^1 dz\,z^N \mathcal{F}^{(1)} (z) & =- {1 \over 4}\zeta_2 + {1\over 4} \int_0^1 dz\,{1-z^{N+1}\over 1-z} \ln(1-z)
\\
& = -{\pi^2\over 16} -\frac18 \left[\psi(N+2)-\psi(1)\right]^2 + \frac18 \psi'(N+2)\,,
\end{align}
where $\psi(x)=d\ln \Gamma(x)/dx$. On the other hand, the expression \re{weakcouplCHF} leads to divergent moments due to the singularity at $z=1$.

Notice that at large $N$, the moments \re{mom} have the double-logarithmic (Sudakov) form $ -(\ln N)^2/8$. This asymptotics is related to the singular behavior of the one-loop correction
to the event shape function at the edge of the phase space, $\mathcal{F}^{(1)} (z)\sim -\ln(1-z)/(1-z)$ for $z\to 1$. Going to higher orders in the weak coupling expansion, we find that the
perturbative corrections to the event shape function get enhanced by powers of $a\ln(1-z)$ and become even more singular as $z\to 1$. It is well known however that such corrections
can be resummed to all orders in the coupling, resulting in event shape functions regular for $z\to 1$ \cite{Kunszt:1989km,Catani:1992ua,Korchemsky:1994is,Dokshitzer:1999sh}.

\section{Weighted cross section from amplitudes}
\label{WeightedAmplitudes}

In this appendix we reproduce the double-scalar correlation using amplitudes. Let us first compute the Born contribution by inserting \re{scattamplx0} into \re{ampweight0},
\begin{align}\label{leadingW}
\vev{ {\cal O} (n) {\cal O}(n') }_{\bf 105}
&
=  \pi \int   \text{dPS}_{2}  (k_1^0\, k_2^0)^{-1} \,  \delta^{(2)}(\Omega_{\vec k_1}-\Omega_{\vec n})\delta^{(2)}(\Omega_{\vec k_2}-\Omega_{\vec n'})
\nonumber\\
&= {1 \over 16 \pi} \int d k_1^0 \, d k_2^0 \, \delta^{(4)} \left(q - k_1^0 n - k_2^0 n'\right)
\, ,
\end{align}
where in the second relation we used the two delta functions to perform the integration over the solid angles $\Omega_{\vec k_1}$ and $\Omega_{\vec k_2}$.
Using the Sudakov decomposition \re{Sudak} in the remaining delta function, $\delta^{(4)} \left(q - k_1^0 n - k_2^0 n'\right) =
\delta^{(2)} (\vec{q}_\perp) \delta \left(k_1^0 - (qn')/(nn')\right) \delta \left(k_2^0 - (qn)/(nn') \right)/$ $(n n')$, the integrals over the energies are
easily computed yielding
\begin{align}
\vev{ {\cal O} (n) {\cal O}(n') }_{\bf 105}
=
{1 \over 16 \pi} { \delta^{(2)}(\vec q_{\perp}) \over (nn')} =  {1 \over (4 \pi)^2} {\delta(1-z) \over q^2 (nn')}
\, .
\end{align}
We now turn to the $O(a)$ correction to $\vev{ {\cal O} (n) {\cal O} (n') }$, Eq.\ \re{ampweight}, away from the kinematical boundaries, for  $0 < z< 1$.
Taking into account \re{scattamplx1} we find from \re{ampweight}
\begin{align}\label{subleadingW}
\vev{ {\cal O} (n) {\cal O}(n') }_{\bf 105}
=  16 \pi^3 a \int   \text{dPS}_{3} {s_{12} \over s_{13} s_{23}}  (k_1^0\, k_2^0)^{-1} \,
\delta^{(2)}(\Omega_{\vec k_1}-\Omega_{\vec n})\delta^{(2)}(\Omega_{\vec k_2}-\Omega_{\vec n'})
\, .
\end{align}
First we can recast the phase-space measure in the form
\begin{align}\label{subleadingWA}
&\int \text{dPS}_{3}  (k_1^0\, k_2^0)^{-1} \,  \delta^{(2)}(\Omega_{\vec k_1}-\Omega_{\vec n})\delta^{(2)}(\Omega_{\vec k_2}-\Omega_{\vec n'})
\\
&\qquad
= {1 \over 4 (2 \pi)^5} \int d^4 k_1 d^4 k_2\int_0^\infty d\tau_1 d\tau_2 \,\delta^{(4)}(k_1-n_1 \tau_1)  \delta^{(4)}(k_2-n_2 \tau_2) \delta_{+}((q-k_1-k_2)^2)
\, .
\nonumber
\end{align}
Then we notice that on the support of the delta functions, the square of the three-particle form factor is independent of the integration variables
(we recall that $s_{ij}=(k_i+k_j)^2$ and $k_1+k_2+k_3=q$),
\begin{align}
{s_{12} \over s_{13} s_{23}} = \frac{(n_1 n_2)}{2(q n_1) (q n_2) - q^2 (n_1 n_2)}\,,
\end{align}
so that we get for the double scalar correlation
\begin{align}
\vev{ {\cal O} (n) {\cal O}(n') }_{\bf 105} =   {a \over 2} {1 \over (2 \pi)^2} \frac{(n_1 n_2)}{2(q n_1) (q n_2) - q^2 (n_1 n_2)}
&
\int_0^\infty d\tau_1 d\tau_2 \delta_{+}((q-\tau_1 n_1-\tau_2 n_2)^2) \nonumber\\
&
= -  { a \over 4} {1 \over (2 \pi)^2} \frac{1}{q^2 (n_1 n_2)} {z \ln(1-z) \over 1 - z}
\, .
\end{align}
This result coincides with the correlation function computation, see Eq.\ \re{weakcouplCHF}.

\section{Mellin representation for $D$-functions}
\label{Dfunctions}

In this Appendix we work out the Mellin representation of the $D$-functions arising in the supergravity calculation of the four-point function of half-BPS operators \re{Strong4point}.
A similar, but different in the details, computation was presented in Ref.\ \cite{Arutyunov:2000ku}. These functions are given by a product of five-dimensional bulk-to-boundary
propagators in anti-de Sitter space and they can be written as  a Schwinger integral,
\begin{align}\label{Sch}
D_{\Delta_1 \Delta_2 \Delta_3 \Delta_4}
=
2 K_{\Delta_1 \Delta_2 \Delta_3 \Delta_4}
\int_0^\infty \prod_{k=1}^4 d t_k t_k^{\Delta_k - 1}  \exp\left( - \sum_{i<j} t_i t_j x_{ij}^2 \right)
\, ,
\end{align}
with the overall normalization chosen as
\begin{align}
K_{\Delta_1 \Delta_2 \Delta_3 \Delta_4}
=
\frac{\pi^2}{2} \frac{\Gamma (\ft12 (\Delta_1 + \Delta_2 + \Delta_3 + \Delta_4) - 2)}{\Gamma (\Delta_1)\Gamma (\Delta_2)\Gamma (\Delta_3)\Gamma (\Delta_4)}
\, .
\end{align}
The representation \re{Sch} is particularly convenient for  rewriting the integral in Mellin form. This is easily achieved by making use of the Symanzik `star formula'
\cite{Symanzik:1972wj},
\begin{align}
\label{SymanzikForm}
\int_0^\infty \prod_{k=1}^4 d t_k t_k^{\Delta_k - 1}  \exp\left( - \sum_{i<j} t_i t_j x_{ij}^2 \right)
=
\frac{1}{2} \int \prod_{i<j}^4 \frac{d \delta_{ij}}{2 \pi i} \, \Gamma ({-}\delta_{ij}) (x_{ij}^2)^{\delta_{ij}}
\, ,
\end{align}
where the integration variables on the right-hand side are not independent and are subject to the constraints
\begin{align}
\delta_{ij} = \delta_{ji}
\, , \qquad
\sum_{j \neq i}^4 \delta_{ij} = -\Delta_i
\, .
\end{align}
Out of the six $\delta$-variables, there are just two independent ones that define the integration measure in the above formula. These can be conveniently chosen
as $j_1 = - \delta_{12}$ and $j_2 = - \delta_{23}$, the rest arising from the conditions relating them to the $\Delta$'s. Using the above formulas for the
$D$-functions, we can accommodate all accompanying prefactors by a shift of the integration variables $j_1$ and $j_2$, such that the Mellin representation
for Eq.\ \re{Strong4point} takes the form \re{strcoupla}.

\end{document}